\documentclass[conference]{IEEEtran}
\IEEEoverridecommandlockouts

\usepackage{mathptmx} 
\newcommand{\ignore}[1]{}
\usepackage[pass]{geometry}
\usepackage[hyphens]{url}
\usepackage{color}
\usepackage{soul}

\def\BibTeX{{\rm B\kern-.05em{\sc i\kern-.025em b}\kern-.08em
    T\kern-.1667em\lower.7ex\hbox{E}\kern-.125emX}}

\usepackage{fancyhdr}
\usepackage[normalem]{ulem}
\usepackage{cite}
\usepackage{epsfig}
\usepackage{dblfloatfix}    
\usepackage{graphicx}
\usepackage{textcomp}
\usepackage{xcolor}
\usepackage{epstopdf}
\usepackage{setspace} 
\usepackage{subfig}
\usepackage{algorithmic}
\usepackage{url}
\usepackage{amsmath,amssymb,amsfonts}
\usepackage{pifont}
\usepackage{fixltx2e}
\usepackage{anyfontsize}
\usepackage{kantlipsum}
\usepackage{multirow}
\usepackage[font=bf,skip=2pt]{caption}
\usepackage{array}
\usepackage{comment}
\usepackage{mwe}
\usepackage{enumitem}

\usepackage{tikz}
\usepackage{kantlipsum}
\newcommand*\blkcircled[1]{\tikz[baseline=(char.base)]{
            \node[shape=circle,draw=red!10!black,fill=white,font=\bfseries,inner sep=1pt] (char) {\textcolor{black}{#1}};}}
\newcommand*\whitecircled[1]{\tikz[baseline=(char.base)]{
            \node[shape=circle,fill=black,font=\bfseries,inner sep=1pt] (char) {\textcolor{white}{#1}};}}

\newcommand{\mycomment}[1]{}

\usepackage{color}
\usepackage{soul}
\usepackage{xspace}

\newcommand{\newedit}[1]{#1\xspace}

\begin{document}
\title{ZnG: Architecting GPU Multi-Processors with New Flash for Scalable Data Analysis}
\author{
{\large Jie Zhang and Myoungsoo Jung} \\
\vspace{-2pt}
       {\large \emph{Computer Architecture and Memory Systems Laboratory,}}\\
\vspace{-2pt}
       {\normalsize{Korea Advanced Institute of Science and Technology (KAIST)}}\\
\vspace{-3pt}
	   {\large {http://camelab.org}}
       }


\maketitle

\begin{abstract}
We propose ZnG, a new GPU-SSD integrated architecture, which can maximize the memory capacity in a GPU and address performance penalties imposed by an SSD. Specifically, ZnG replaces all GPU internal DRAMs with an ultra-low-latency SSD to maximize the GPU memory capacity. ZnG further removes performance bottleneck of the SSD by replacing its flash channels with a high-throughput flash network and integrating SSD firmware in the GPU's MMU to reap the benefits of hardware accelerations. Although flash arrays within the SSD can deliver high accumulated bandwidth, only a small fraction of such bandwidth can be utilized by GPU's memory requests due to mismatches of their access granularity. To address this, ZnG employs a large L2 cache and flash registers to buffer the memory requests. Our evaluation results indicate that ZnG can achieve 7.5$\times$ higher performance than prior work.
\end{abstract}

\begin{IEEEkeywords}
data movement, GPU, SSD, heterogeneous system, MMU, L2 cache, Z-NAND, DRAM
\end{IEEEkeywords}

\section{Introduction}
\label{sec:intro}
Over the past few years, graphics processing units (GPUs) become prevailing to accelerate the large-scale data-intensive applications such as graph analysis and bigdata \cite{jiang2015scaling, napoli2014cloud, li2015heterospark, hong2011efficient, wang2016gunrock}, because of the high computing power brought by their massive cores. To reap the benefits from the GPUs, large-scale applications are decomposed into multiple GPU kernels, each contains ten or hundred of thousands of threads. These threads can be simultaneously executed by such GPU cores, which exhibits high thread-level parallelism (TLP). While the massive parallel computing drives GPUs to exceed CPUs' performance by upto 100 times, the on-board memory capacity of GPUs is much less than that of the host-side main memory, which cannot accommodate all data sets of the large-scale applications. In practice, as a peripheral device, GPUs have the limited on-board space to deploy an enough number of memory packages \cite{gtx2080}, while the capacity of a single memory package is difficult to expand due to the DRAM scaling issues such as DRAM cell disturbances and retention time violations \cite{kim2015architectural, kang2014co, nair2013archshield}.   

To meet the requirement of such large memory capacity, a GPU vendor utilizes the underlying NVMe SSD as a swap disk of the GPU memory and leverages the memory management unit (MMU) in GPUs to realize memory virtualization \cite{SSG}. For example, if a data block requested by a GPU core misses in the GPU memory, GPU's MMU raises the exception of a page fault. As both GPU and NVMe SSD are peripheral devices, the GPU informs the host to service the page fault, which unfortunately introduces severe data movement overhead. 
Specifically, the host first needs to load the target page from the NVMe SSDs to the host-side main memory and then moves the same data from the memory to the GPU memory. The data copy across different computing domains, the limited performance of NVMe SSD and the bandwidth constraints of various hardware interfaces (i.e., PCIe) significantly increase the latency of servicing page faults, which in turn degrades the overall performance of many applications at the user-level.

\begin{figure}
\centering
\subfloat[HybridGPU design.]{\label{fig:flashgpu-arch1}\rotatebox{0}{\includegraphics[width=0.44\linewidth]{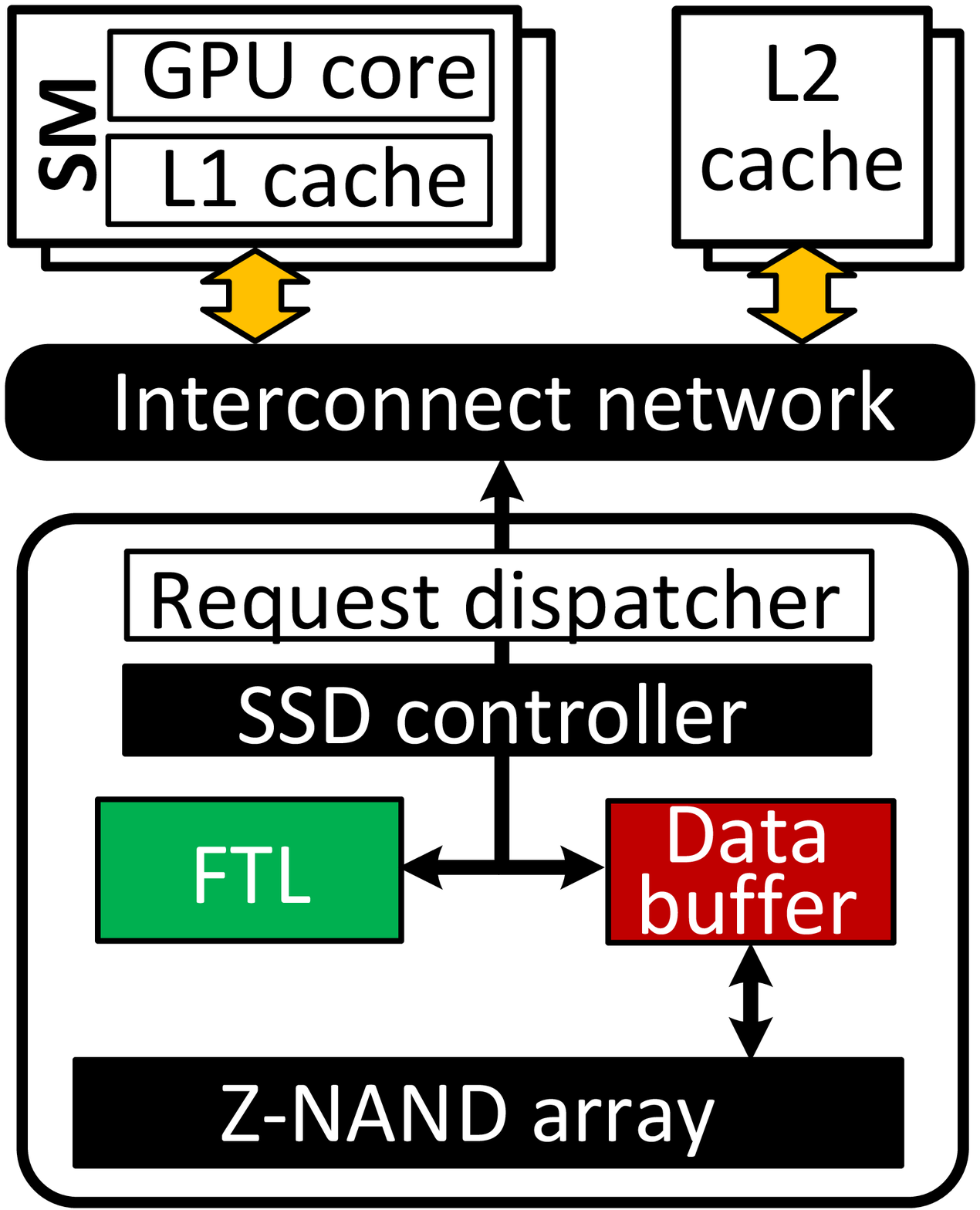}}}
\hspace{4pt}
\subfloat[Bandwidth.]{\label{fig:motiv5-DRAM-flash-1}\rotatebox{0}{\includegraphics[width=0.51\linewidth]{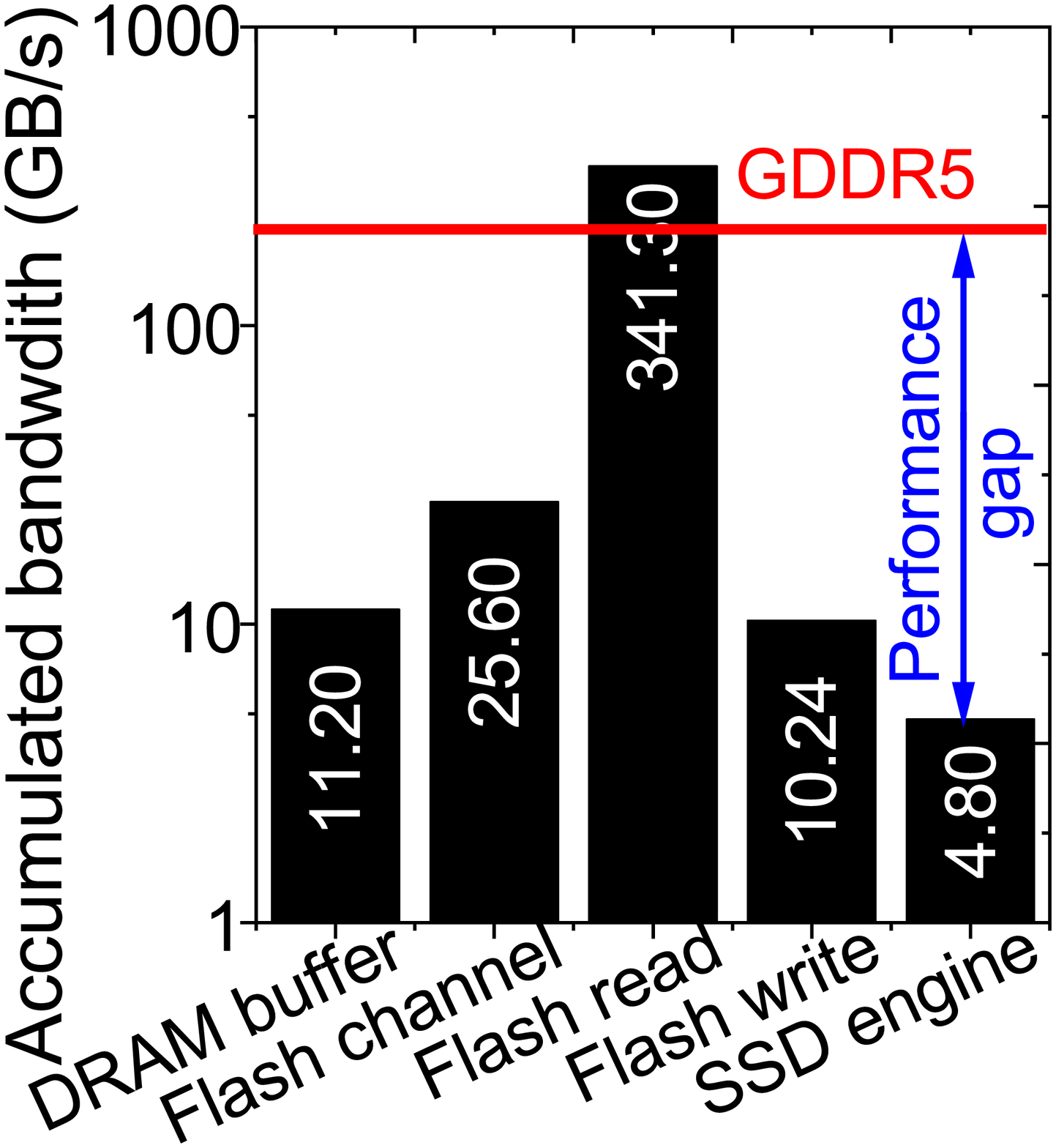}}}
\caption{\label{fig:motiv1}An integrated HybridGPU architecture and the performance analysis.\vspace{-15pt}}
\vspace{-10pt}
\end{figure}

To reduce the data movement overhead, a prior study \cite{zhang2019flashgpu}, referred to as \textit{HybridGPU}, proposes to directly replace a GPU's on-board DRAM packages with Z-NAND flash packages as shown in Figure \ref{fig:flashgpu-arch1}. Z-NAND, as a new type of NAND flash, achieves 64 times higher capacity than DRAM, while reducing the access latency of conventional flash media from hundreds of micro-seconds to a few micro-seconds \cite{koh2018exploring}. However, Z-NAND faces several challenges to service the GPU memory requests directly: 1) the minimum access granularity of Z-NAND is a page, which is not compatible with the memory requests; 2) Z-NAND programming (writes) requires the assistance of SSD firmware to manage address mapping as it forbids in-place updates; and 3) its access latency is still much longer than DRAM. To address these challenges, HybridGPU employs a customized SSD controller to execute SSD firmware and has a small DRAM as read/write buffer to hide the relatively long Z-NAND latency. While this approach can eliminate the data movement overhead by placing Z-NAND close to GPU, there is a huge performance disparity between this approach and the traditional GPU memory subsystem. To figure out the main reason behind the performance disparity, we analyze the bandwidths of traditional GPU memory subsystem and each component in HybridGPU. The results are shown in Figure \ref{fig:motiv5-DRAM-flash-1}. The maximum bandwidth of HybridGPU's internal DRAM buffer is 96\% lower than that of the traditional GPU memory subsystem. This is because the state-of-the-art GPUs employ six memory controllers to communicate with a dozen of DRAM packages via a 384-bit data bus \cite{nvidia2018nvidia}, while the DRAM buffer is a single package connected to a 32-bit data bus \cite{zhang2019flashgpu}. It is possible to increase the number of DRAM packages in HybridGPU by reducing the number of its Z-NAND packages. However, this solution is undesirable as it can reduce the total memory capacity in GPU. In addition, I/O bandwidth of the flash channels and data processing bandwidth of a SSD controller are much lower than those of the traditional GPU memory subsystem, which can also become performance bottleneck. This is because the bus structure of the flash channels constrains itself from scaling up with a higher frequency, while the single SSD controller has a limited computing power to translate the addresses of the memory requests from all the GPU cores. 

We propose a new GPU-SSD architecture, \textit{ZnG}, to maximize the memory capacity in the GPU and to address the performance penalties imposed by the SSD integrated in the GPU. ZnG replaces all the GPU on-board memory modules with multiple Z-NAND flash packages and directly exposes the Z-NAND flash packages to the GPU L2 cache to reap the benefits of the accumulated flash bandwidth. Specifically, to prevent the single SSD controller from blocking the services of memory requests, ZnG removes HybridGPU's request dispatcher. Instead, ZnG directly attaches the underlying flash controller to a GPU interconnect network to directly adopt memory requests from GPU L2 caches. Since the SSD controller also has a limited computing power to perform address translation for the massive number of memory requests, ZnG offloads the functionality of address translation to the GPU internal MMU and flash address decoder (to hide the computation overhead). To satisfy the accumulated bandwidth of Z-NAND flash arrays, ZnG also increases the bandwidth of the flash channels. Lastly, although replacing all GPU DRAM modules with the Z-NAND packages can maximize the accumulated flash bandwidth, such bandwidth can be underutilized without a DRAM buffer, as Z-NAND has to fetch a whole 4KB flash page to serve a 128B data block. ZnG increases the GPU L2 cache sizes to buffer the pages fetched from Z-NAND, which can serve the read requests, while it leverages Z-NAND internal cache registers to accommodate the write requests. Our evaluation results indicate that ZnG can improve the overall GPU performance by 7.5$\times$, compared to HybridGPU.
Our \textbf{contributions} of this work can be summarized as follows:

\noindent $\bullet$ \emph{New GPU-SSD architecture to remove performance bottleneck.}
The HybridGPU's request dispatcher can be bottleneck to interact with both the underlying flash controllers and all L2 cache banks. To address this, ZnG directly attaches multiple flash controllers to the GPU interconnect network, such that the memory requests, generated by GPU L2 caches, can be served across different flash controllers in an interleaved manner. On the other hand, compared to the GPU interconnect network, a flash channel has a limited number of electrical lanes and runs in a relatively low clock frequency, and therefore, its bandwidth is much lower than the accumulated bandwidth of the Z-NAND arrays. ZnG addresses this by changing the flash network from a bus to a mesh structure.

\noindent $\bullet$ \emph{Hardware implementation for zero-overhead address translation.}
The large-scale data analysis applications are typically read-intensive. By leveraging characteristics of the target user applications, ZnG splits the flash address translation into two parts. First, a read-only mapping table is integrated in the GPU internal MMU and cached by the GPU TLB. All memory requests can directly get their physical addresses when MMU looks up the mapping table to translate their virtual addresses. Second, when there is a memory write, the data and the updated address mapping information are simultaneously recorded in the flash address decoder and flash arrays. A future access to the written data will be remapped by the flash address decoder. 

\noindent $\bullet$ \emph{L2 cache and flash internal register design for maximum flash bandwidth.}
As Z-NAND is directly connected to GPU L2 cache via flash controllers, ZnG increases the L2 cache capacity with an emerging non-volatile memory, STT-MRAM, to buffer more number of pages from Z-NAND. ZnG further improves the space utilization of the GPU L2 caches by predicting spatial locality of the fetched pages. As STT-MRAM suffers from a long write latency, ZnG constructs L2 cache as a read-only cache. To accommodate the write requests, ZnG increases the number of registers in Z-NAND flash and configures all the registers within a same flash package as fully-associative cache to accommodate requests.

\section{Background}
\label{sec:background}
\begin{figure}
	\centering
	\includegraphics[width=1\linewidth]{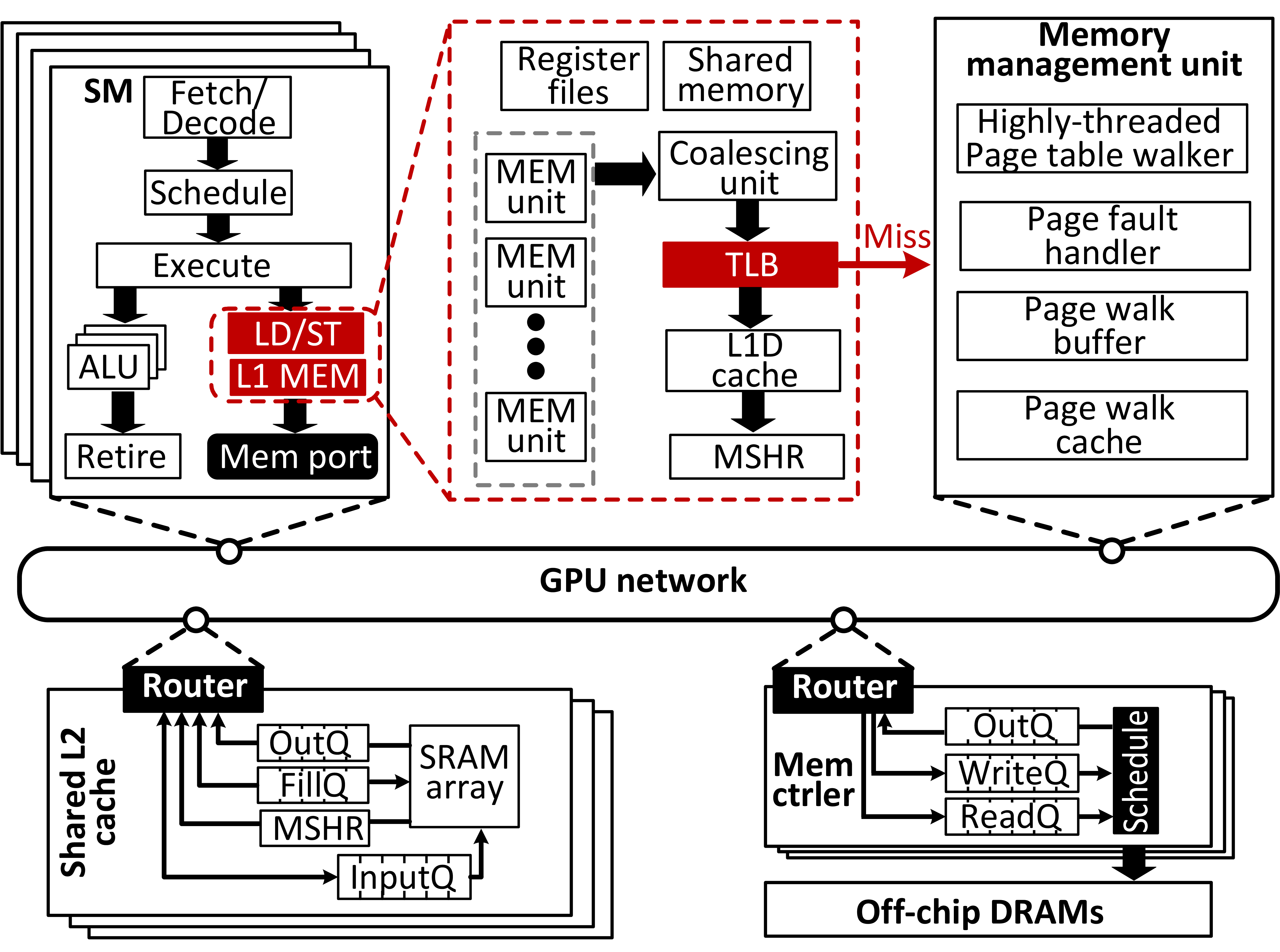}
	\vspace{-10pt}
	\caption{\label{fig:gpu-arch}GPU internal architecture.\vspace{-5pt}}
	\vspace{-15pt}
\end{figure}


\subsection{GPU internal architecture}
Figure \ref{fig:gpu-arch} shows a representative GPU architecture, in which streaming multiprocessors (SMs), shared L2 caches, memory controllers and a memory management unit are connected via a GPU internal network. 
Within SMs, multiple arithmetic logic units (ALU) are employed to execute a group of 32 threads, called \emph{warp} \cite{nvidia2009nvidia}, in a lock step. 
During the execution, the instructions of each warp are fetched and decoded first. The warp scheduler then schedules an available warp and issues the decoded instructions of this warp. Based on the instruction types, the arithmetic instructions are executed in ALUs, while the load/store instructions access the on-chip memory via the LD/ST units. 
The on-chip memory is comprised of L1D cache and shared memory. Before accessing the L1D cache, the memory requests generated by thirty-two threads in a warp are issued to the coalescing unit. This logic unit exams the request addresses and merges the 32-bit memory requests into fewer but larger memory requests to improve the utilization of L1D cache bandwidth. Note that the state-of-the-art GPUs integrates a translation lookaside buffer (TLB)/memory management unit (MMU) to support memory virtualization \cite{nvidia2018nvidia, armgpummu}. Before accessing the L1D cache, the virtual addresses of memory requests need to be translated to the logical addresses via TLB and MMU \cite{mps}. Afterwards, if L1D cache hits, the requests can be served from the L1D cache. Otherwise, the requests are tracked by miss status holding registers (MSHR) and forwarded to L2 cache \cite{nvidia2012nvidia}. 

\begin{figure}
\centering
\subfloat[Memory package capacity.]{\label{fig:density}\rotatebox{0}{\includegraphics[width=0.48\linewidth]{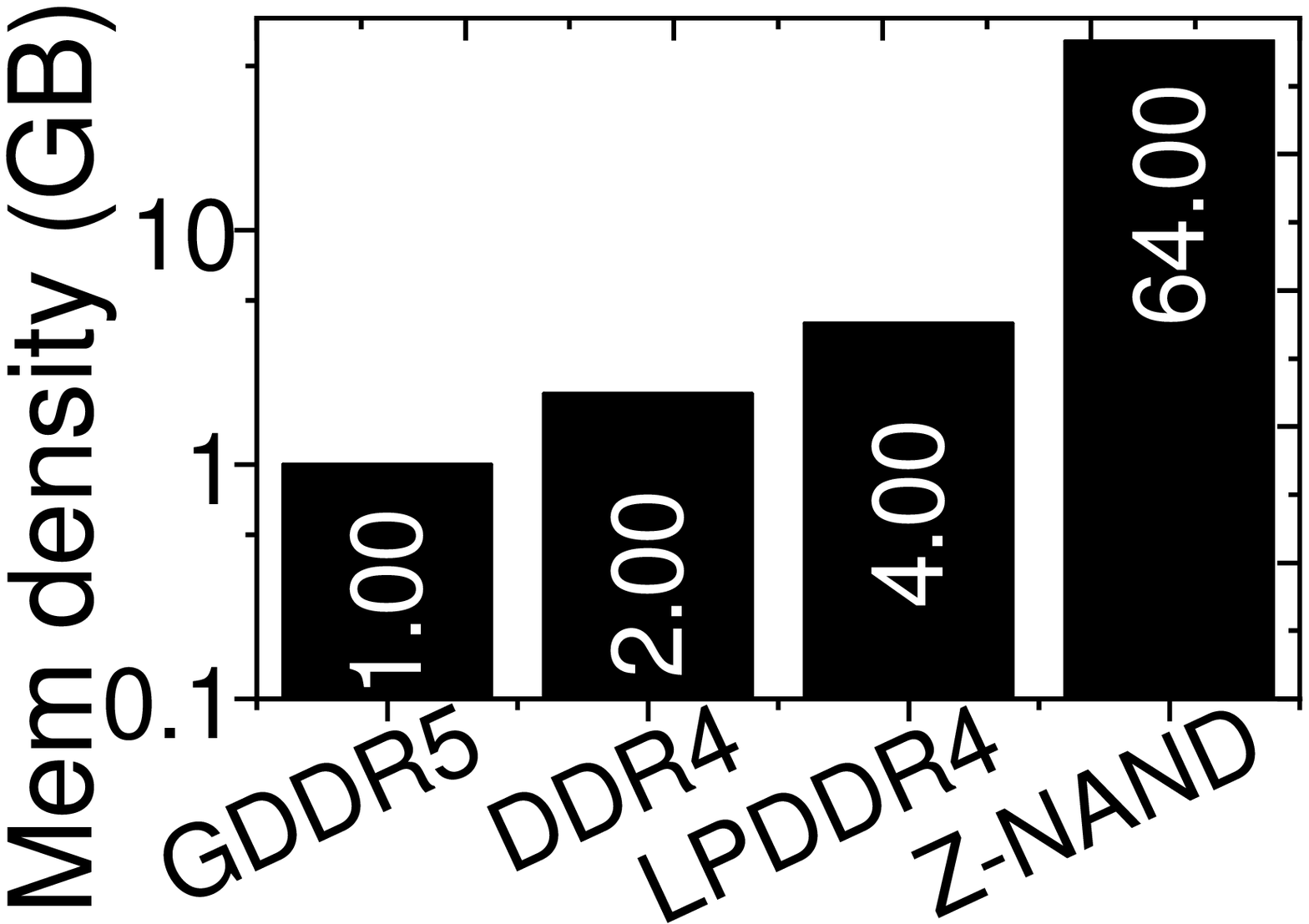}}}
\hspace{3pt}
\subfloat[Power consumption.]{\label{fig:power}\rotatebox{0}{\includegraphics[width=0.48\linewidth]{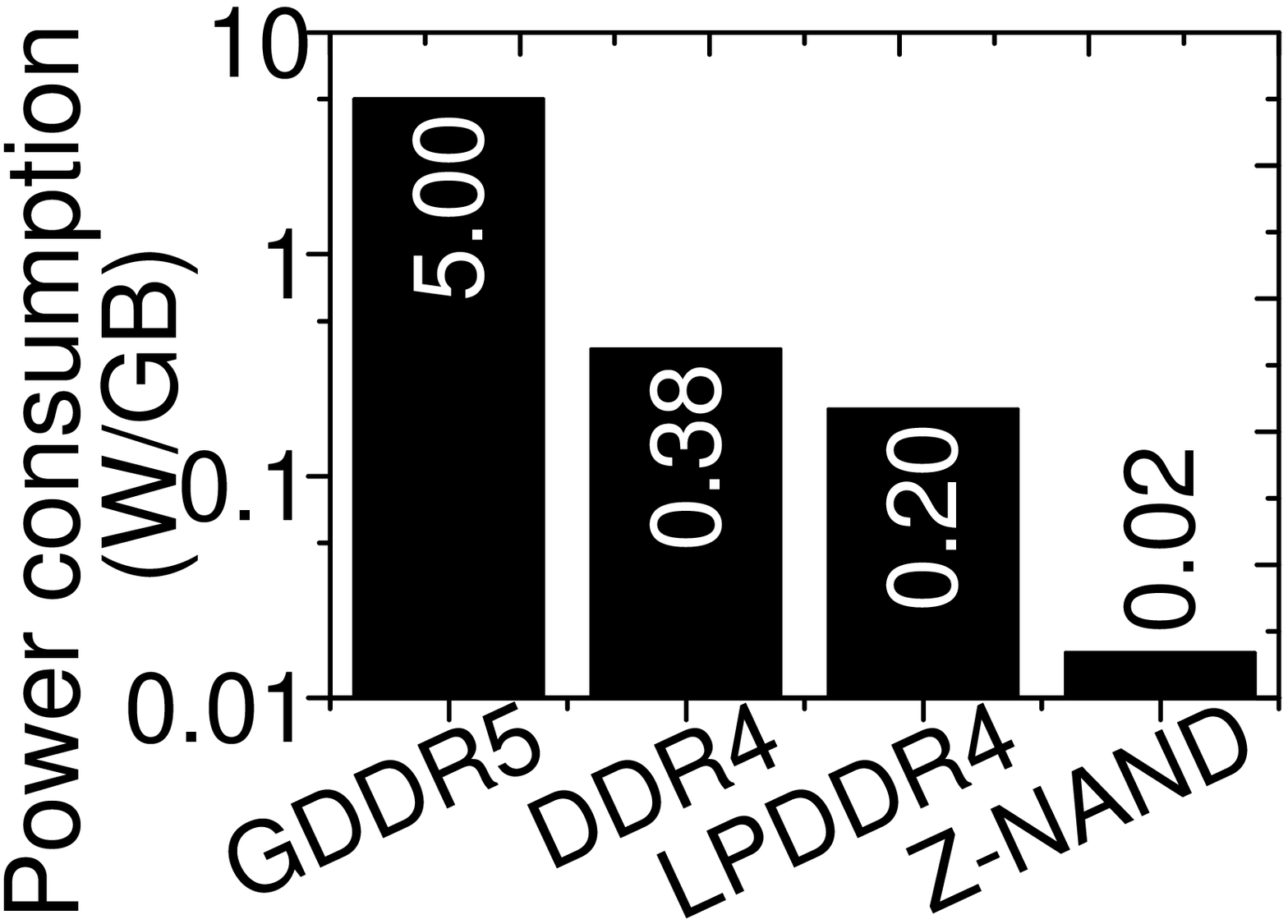}}}
\caption{\label{fig:motiv1}Density and power consumption analysis.\vspace{-5pt}}
\vspace{-20pt}
\end{figure}

Since NVIDIA, AMD and ARM have not documented their GPU memory virtualization designs publicly, we adopt the virtual memory design from academia in this work \cite{power2014supporting}. As shown in Figure \ref{fig:gpu-arch}, the MMU is a shared resource for all SMs, including a highly-threaded page table walker, page walk buffer, page fault handler and page walk cache. The page table walker includes a hardware state machine to walk the page table and a set of page walk buffers. For each L1D TLB miss, the hardware state machine allocates a page walk buffer entry, which records the corresponding address and state. It then issues the memory request associated with the buffer entry and waits for the request completion. To improve the throughput of address translation, the page table walker employs 32 threads. Since the memory accesses cost hundreds of cycles \cite{bakhoda2009analyzing}, GPU MMU also employs a page walk cache to reduce the number of memory accesses. In addition, as a GPU has a limited memory capacity to accommodate all the working sets of multiple applications, GPU MMU employs a page fault handling logic. Specifically, the page fault handler programs page fault information in CPU-side hardware register and sets an interrupt to the CPU. The CPU then copies the target data to the GPU memory in response to the interrupt. 

To achieve high data access bandwidth, GPUs usually employ a shared L2 cache and high-performance off-chip memory. The shared L2 cache exhibits a larger space than the L1D cache, and it is partitioned into multiple tiles to support parallel data accesses. Once the memory requests miss in L2 cache, they are forwarded to the GPU memory. To process the massive memory requests, GPUs typically employ multiple memory controllers (i.e., 6$\sim$8 \cite{nvidia2012nvidia, nvidia2018nvidia}), each connecting to a set of GPU DRAMs to deliver a high bandwidth.

\begin{figure*}
\centering
\subfloat[SSD internal.]{\label{fig:zssd-internal}\rotatebox{0}{\includegraphics[width=0.35\linewidth]{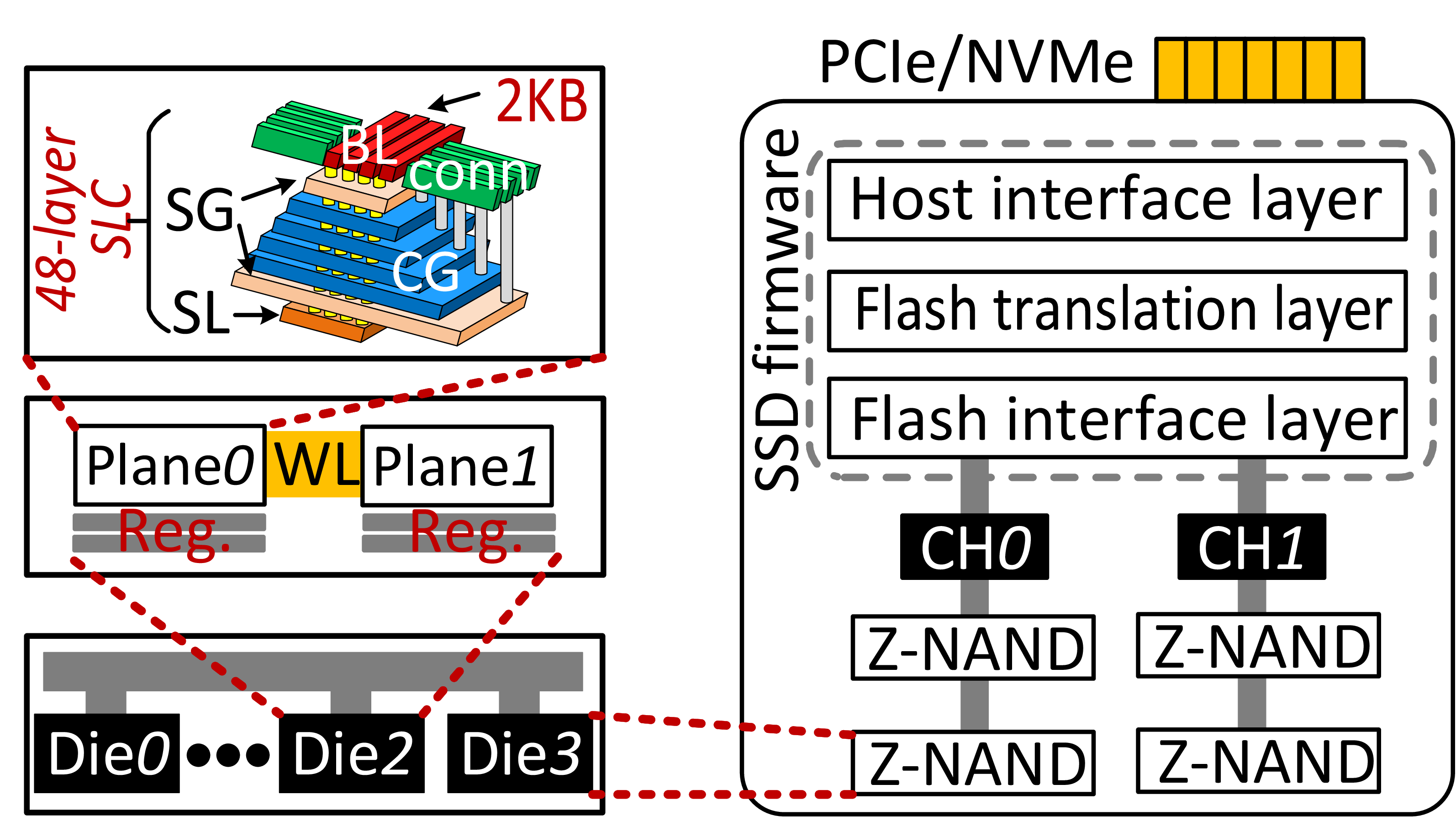}}}
\hspace{1pt}
\subfloat[GPU-SSD architecture.]{\label{fig:hetero-arch}\rotatebox{0}{\includegraphics[width=0.19\linewidth]{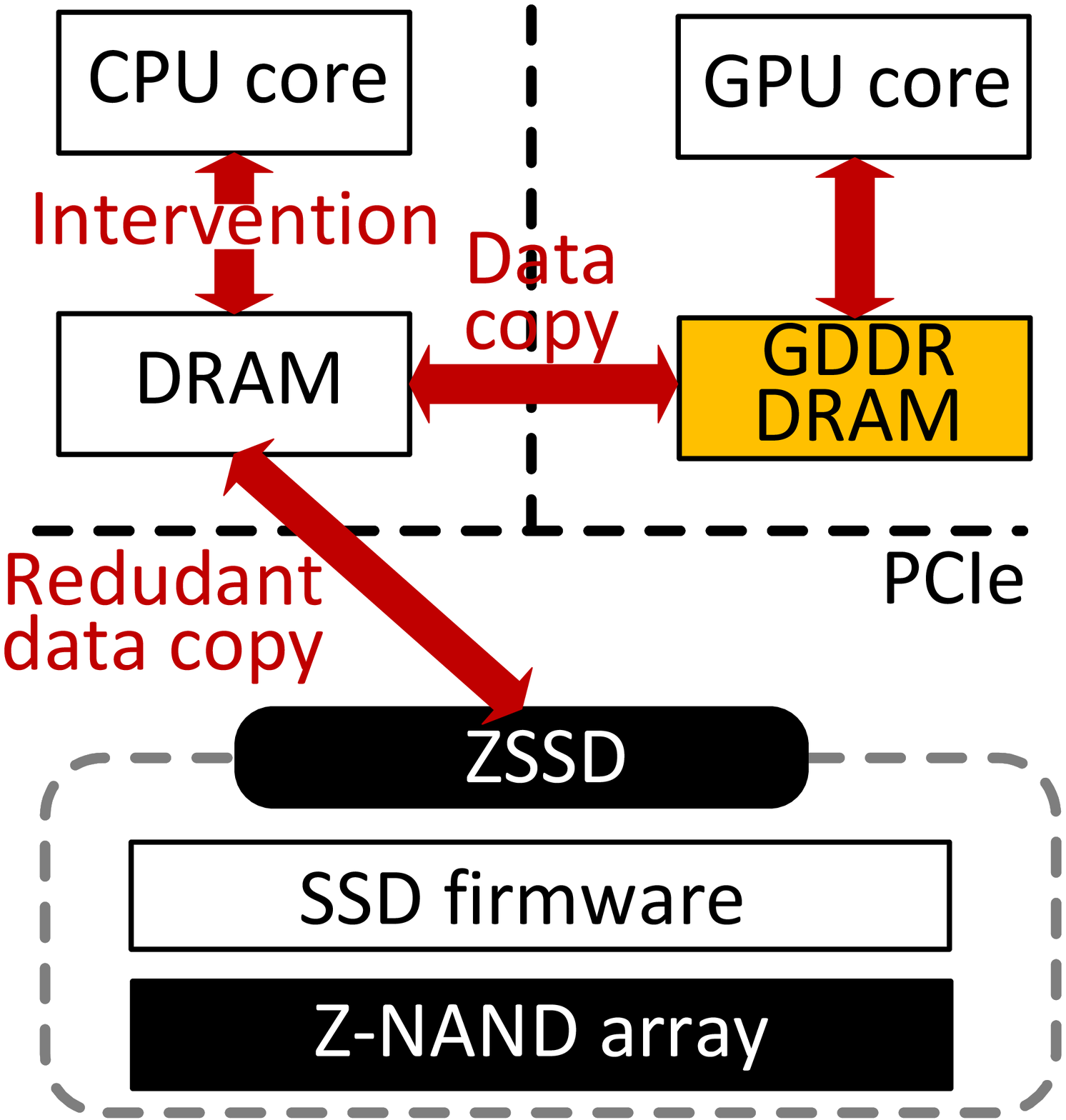}}}
\hspace{1pt}
\subfloat[Performance.]{\label{fig:motiv2-1}\rotatebox{0}{\includegraphics[width=0.16\linewidth]{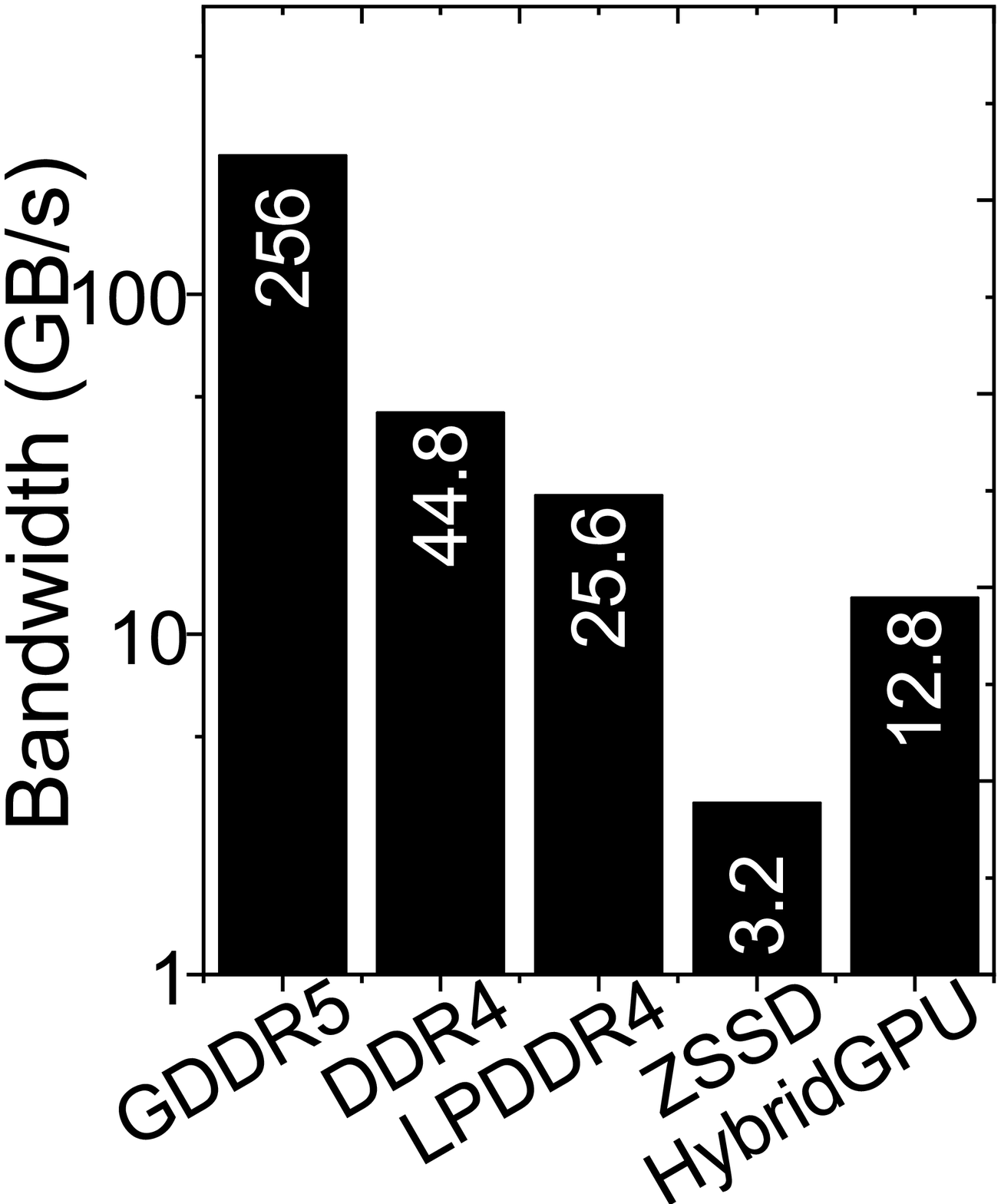}}}
\hspace{1pt}
\subfloat[Latency breakdown.]{\label{fig:motiv7-latency}\rotatebox{0}{\includegraphics[width=0.26\linewidth]{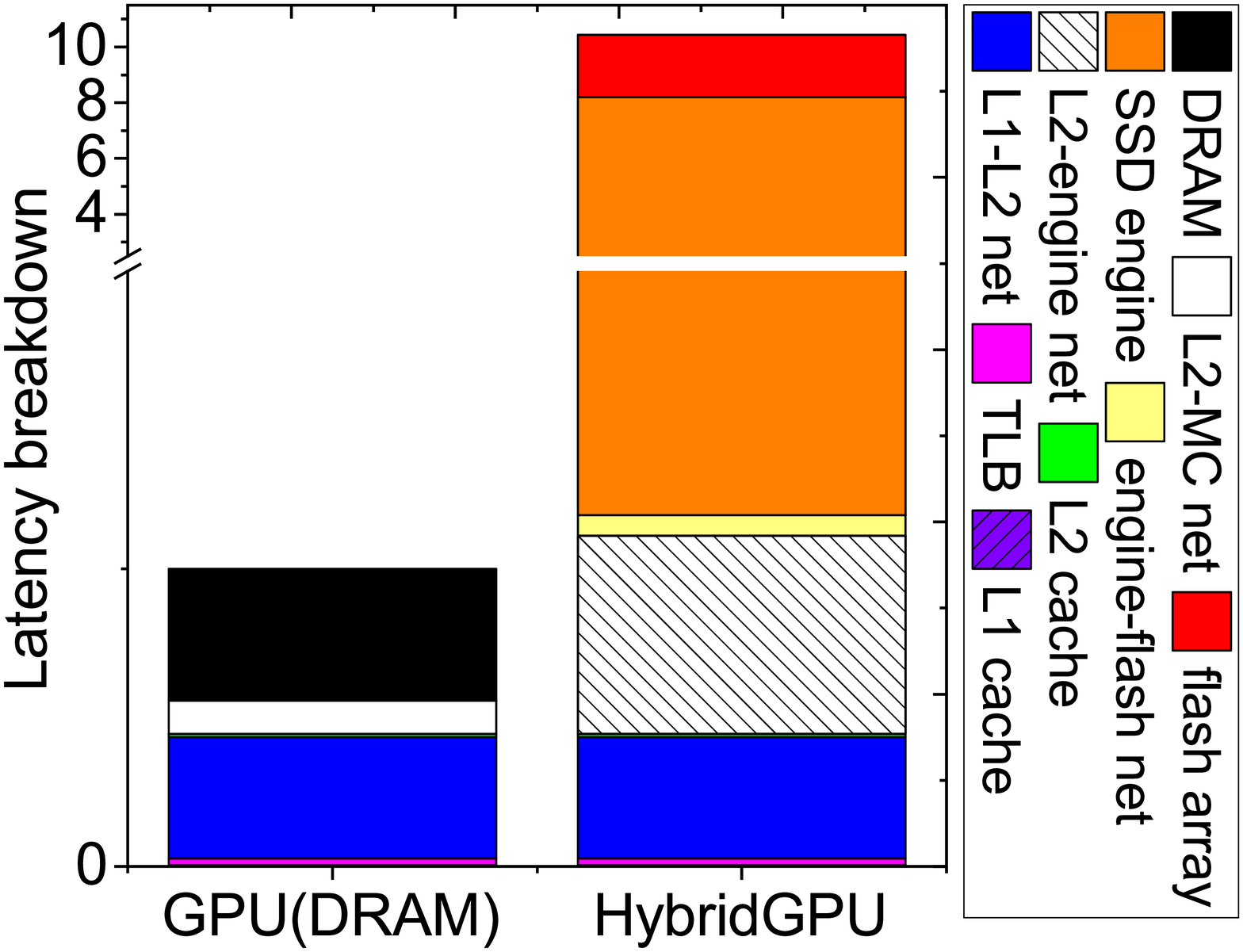}}}
\caption{\label{fig:motiv2}SSD internal architecture, GPU-SSD architecture and the throughput of different memory media.\vspace{-5pt}}
\vspace{-10pt}
\end{figure*}

\subsection{The new NAND flash}
While GPU DRAM provides promising throughput, it has two drawbacks: \emph{low capacity} and \emph{high power consumption}. Figures \ref{fig:density} and \ref{fig:power} compare GPU DRAM (GDDR5), Desktop DRAM (DDR4), mobile DRAM (LPDDR4) and the new NAND flash (\emph{Z-NAND}) in terms of memory density and power consumption (watt per GB), respectively. One can observe from the figures that GPU DRAM has much lower memory density and spends much higher power consumption compared to other types of DRAM and Z-NAND. Even though industry successfully improves the power efficiency of DRAM (i.e., LPDDR4), it is unable to increase the DRAM capacity, which turns out 64$\times$ lower memory density than Z-NAND. 

Compared to DRAM, Z-NAND exhibits the highest memory density and the best power efficiency. In addition, Z-NAND provides a promising performance in terms of both latency and throughput. Figure \ref{fig:zssd-internal} shows the architectural details of Z-NAND. Z-NAND increases its memory density by piling up the flash cells over 48 vertically-stacked layers. Z-NAND further reduces its access latency by reorganizing the low-level flash cell technology and micro-architecture. Specifically, Z-NAND leverages single-level cell (SLC) technologies to shorten flash-level read/program latency and significantly increase its program/erase (P/E) cycles. Thus, read and program operations of Z-NAND take 3us and 100us, respectively, which are 17$\times$ and 6$\times$ faster than those of the state-of-the-art TLC V-NAND flash \cite{koh2018exploring}. \newedit{Z-NAND's P/E cycles (i.e., 100,000) are also higher than those of V-NAND flash (i.e., 3,000$\sim$10,000) by 14$\times$ in overall.} The flash backbone of an SSD that employs Z-NAND can deliver a scalable throughput by fully leveraging its internal parallelism, including the multiple flash channels, packages, dies and planes. 

Unlike GPU DRAM, which is byte-addressable, Z-NAND should serve the reads/writes in a page basis. In addition, Z-NAND forbids overwrites and only allows an in-order programming to write data, due to the page-to-page interference/disturbance. Thus, Z-NAND follows an \emph{erase-before-write} rule \cite{koh2018exploring}; a whole flash block needs to be erased before re-programming its flash cells. 
Because of this erase-before-write rule, modern SSDs employ an embedded CPU and an internal DRAM, referred as \emph{SSD engine}, to execute SSD firmware. The main role of SSD firmware is to translate the logical addresses of incoming I/O requests to the physical address of Z-NAND flash (cf., FTL). When the number of clean flash blocks is under a threshold, SSD firmware performs a clean-up process, called \emph{garbage collection} (GC). Specifically, it migrates multiple valid pages from fully programmed block(s) to clean block(s), erases the target block(s), and updates the mapping information in the internal DRAM.

\subsection{Prior work and motivation}
To address the aforementioned challenges of GPU DRAM, prior work attempts to increase the GPU memory with SSD \cite{zhang2015nvmmu,zhang2019flashgpu}. Figures \ref{fig:hetero-arch} and \ref{fig:flashgpu-arch1} respectively show the overviews of two typical solutions: a GPU-SSD system \cite{zhang2015nvmmu} and a HybridGPU system \cite{zhang2019flashgpu}. The GPU-SSD system employs a discrete GPU and SSD as peripheral devices, which are connected to CPU via a PCIe interface. When page faults occur in GPU due to the limited memory space, CPU serves the page faults by accessing data from the underlying SSD and moving the data to GPU memory through GPU software framework. In addition, the page faults require redundant data copies in the host side due to the user/privilege mode switches. This, unfortunately, wastes host CPU cycles and reduces data access bandwidth. 
In contrast, the HybridGPU directly integrates the SSD into the GPU, which can eliminate CPU intervention and avoid the redundant data copies. While the HybridGPU exhibits much better performance than the GPU-SSD system, memory bandwidth of its embedded SSD module is not comparable to that of GPU memory. Figure \ref{fig:motiv2-1} compares the maximum data access throughput of GPU DRAM, desktop DRAM, mobile DRAM, GPU-SSD and HybridGPU systems. For the GPU-SSD and HybridGPU systems, data are assumed to reside in the SSD. As shown in the figure, GPU DRAM outperforms the GPU-SSD and HybridGPU systems by 80 times and 40 times, respectively. This in turn makes the SSD performance bottleneck in such systems when executing the applications with large-scale data sets. 

\section{High Level View}
\label{sec:highlevelview}

\begin{figure*}
\centering
\subfloat[Performance degradation.]{\label{fig:motiv-baselinevsflashGPU}\rotatebox{0}{\includegraphics[width=0.24\linewidth]{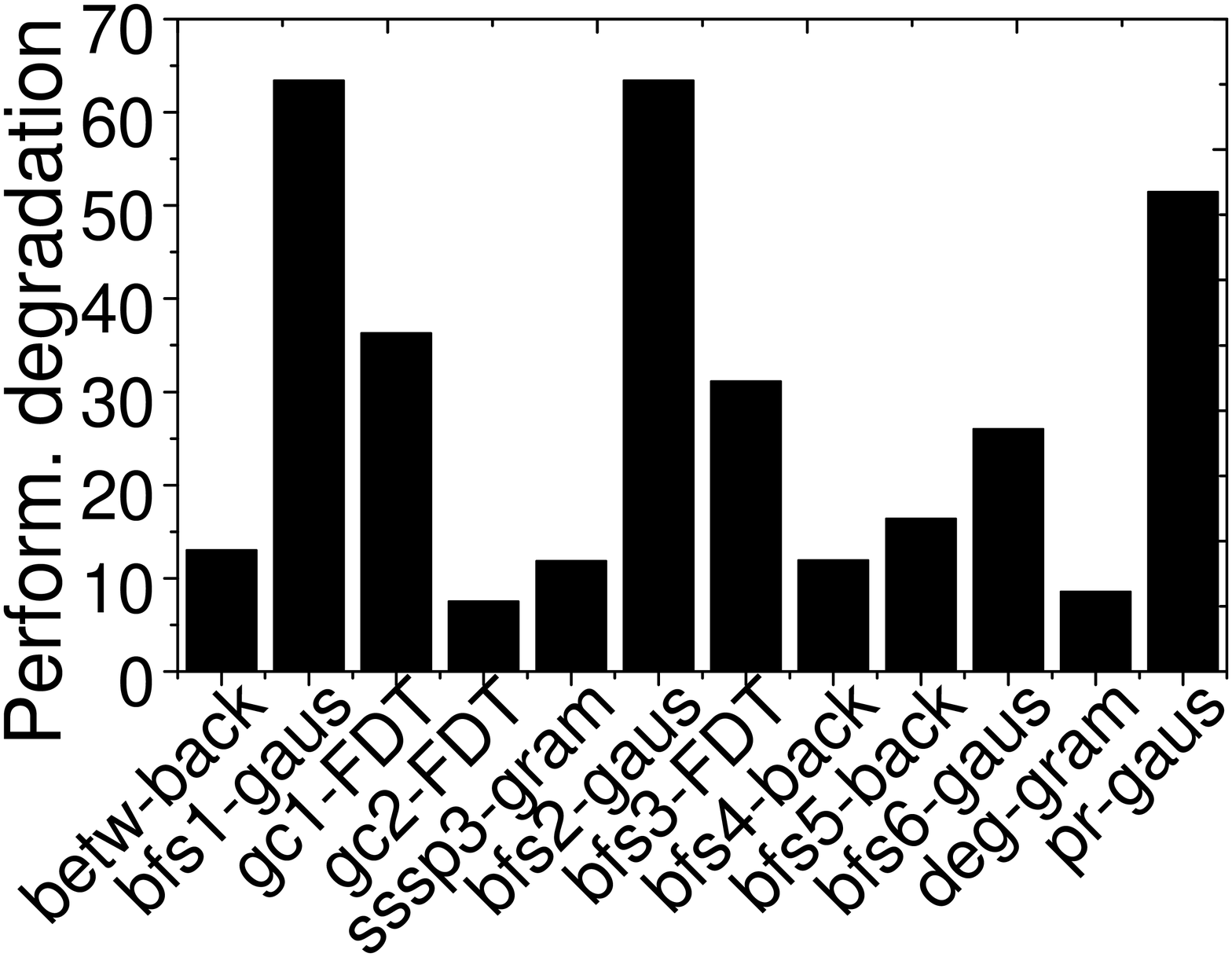}}}
\hspace{1pt}
\subfloat[Read re-accesses.]{\label{fig:motiv-readreaccess}\rotatebox{0}{\includegraphics[width=0.24\linewidth]{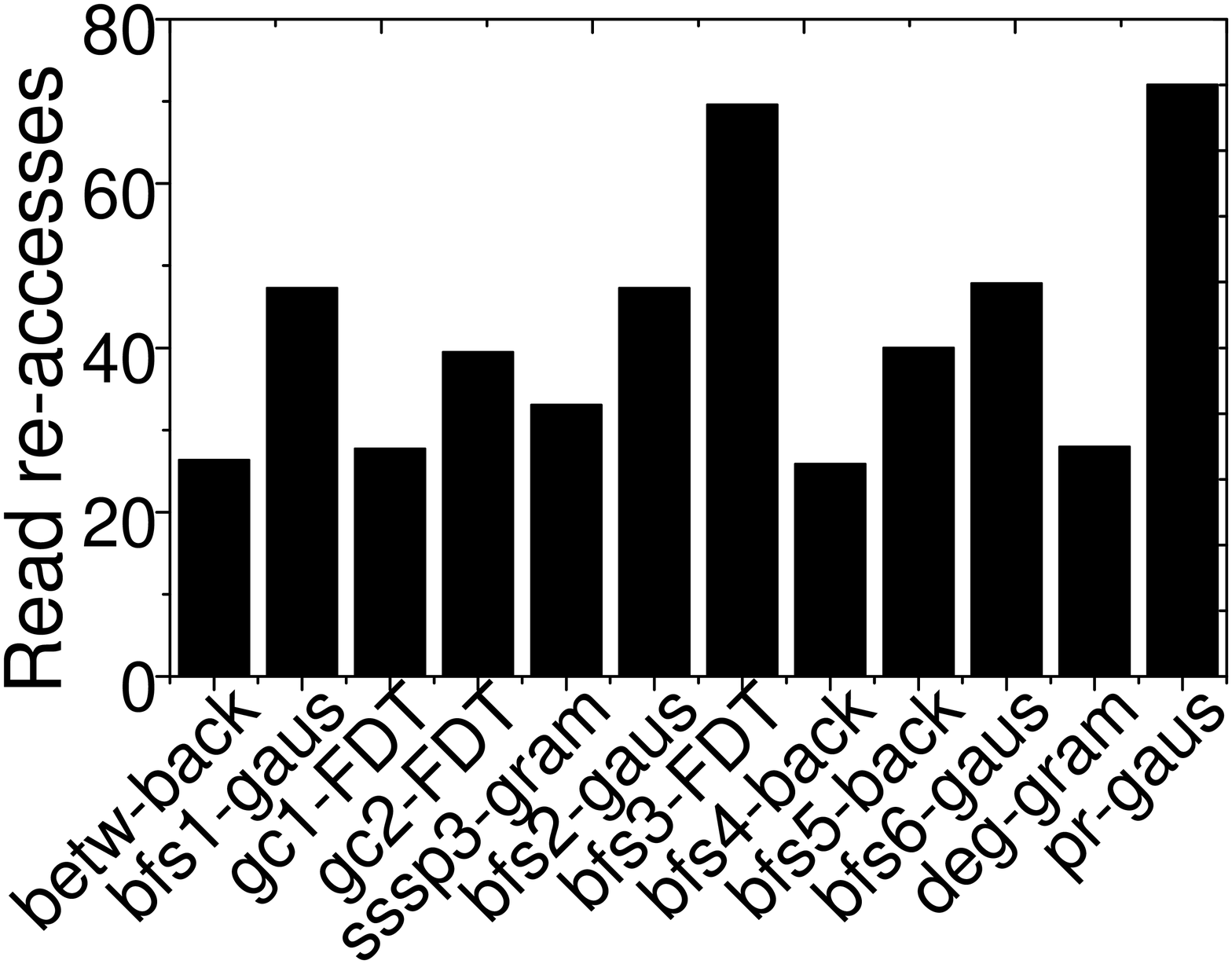}}}
\hspace{1pt}
\subfloat[Write redundancy.]{\label{fig:motiv-writeamp}\rotatebox{0}{\includegraphics[width=0.24\linewidth]{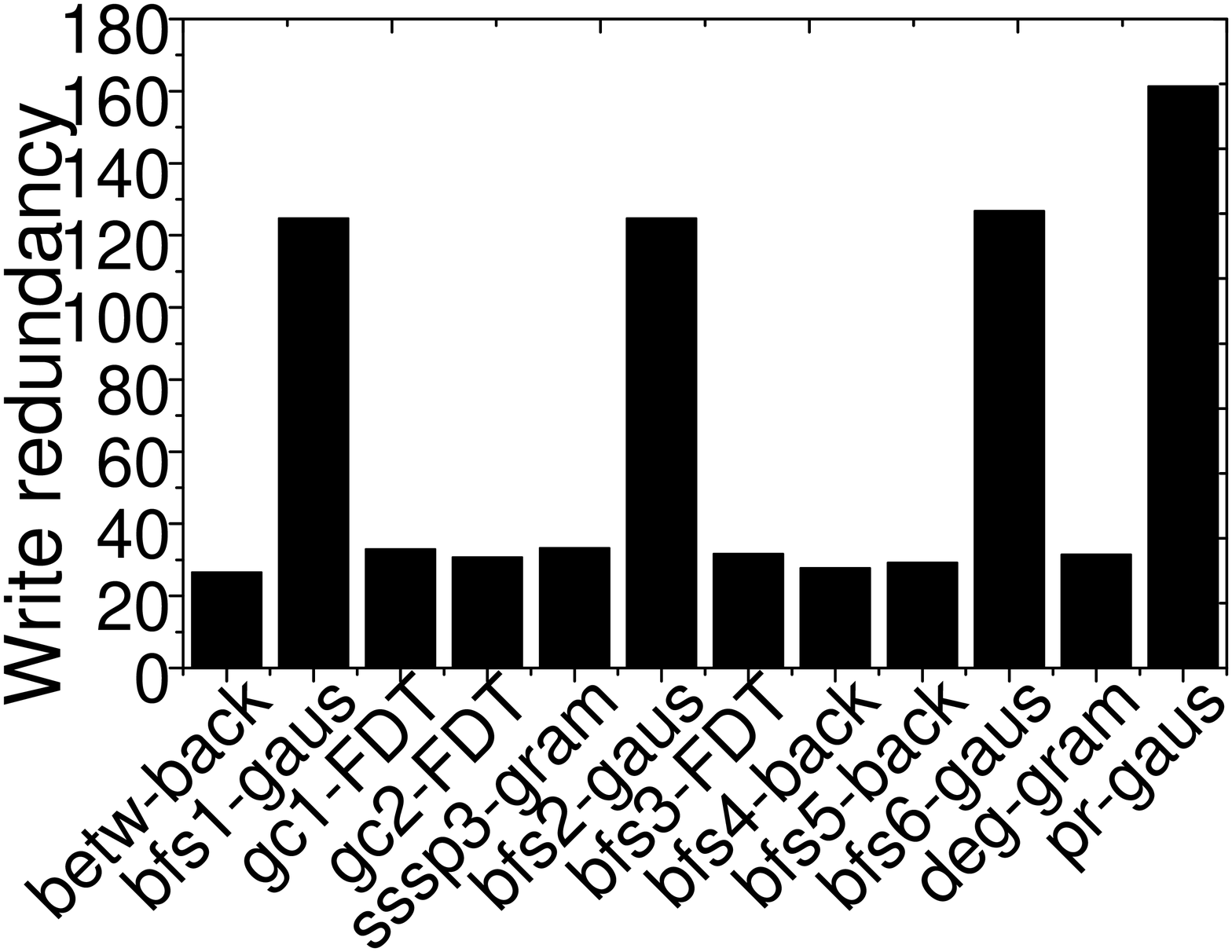}}}
\hspace{1pt}
\subfloat[Access breakdown.]{\label{fig:motiv8-fraction}\rotatebox{0}{\includegraphics[width=0.24\linewidth]{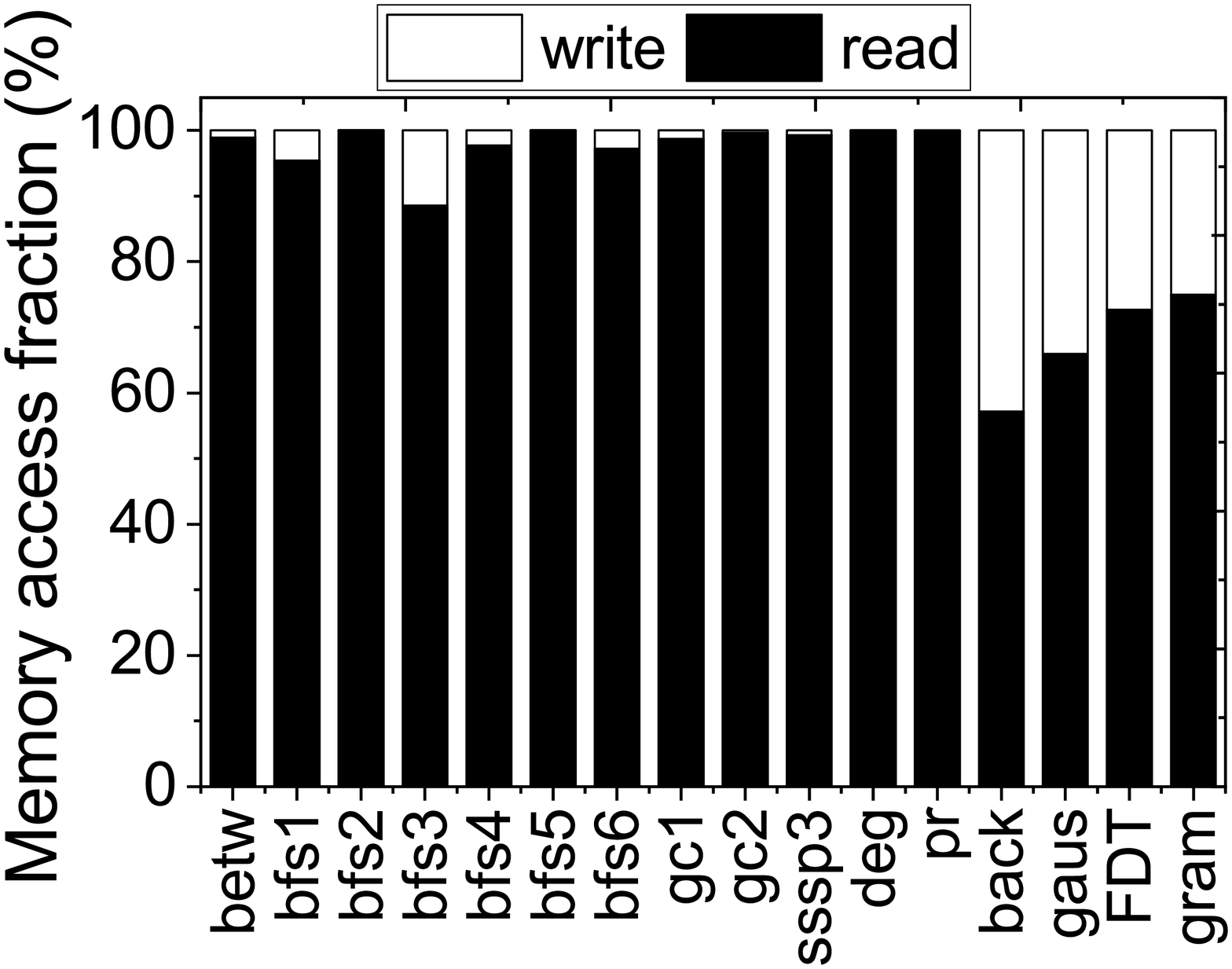}}}
\caption{\label{fig:motiv3}Performance of SSD components, performance degradation, read re-accesses and write redundancy.\vspace{-5pt}}
\vspace{-10pt}
\end{figure*}



\begin{figure*}
\centering
\subfloat[High-level view.]{\label{fig:highlevelview}\rotatebox{0}{\includegraphics[width=0.26\linewidth]{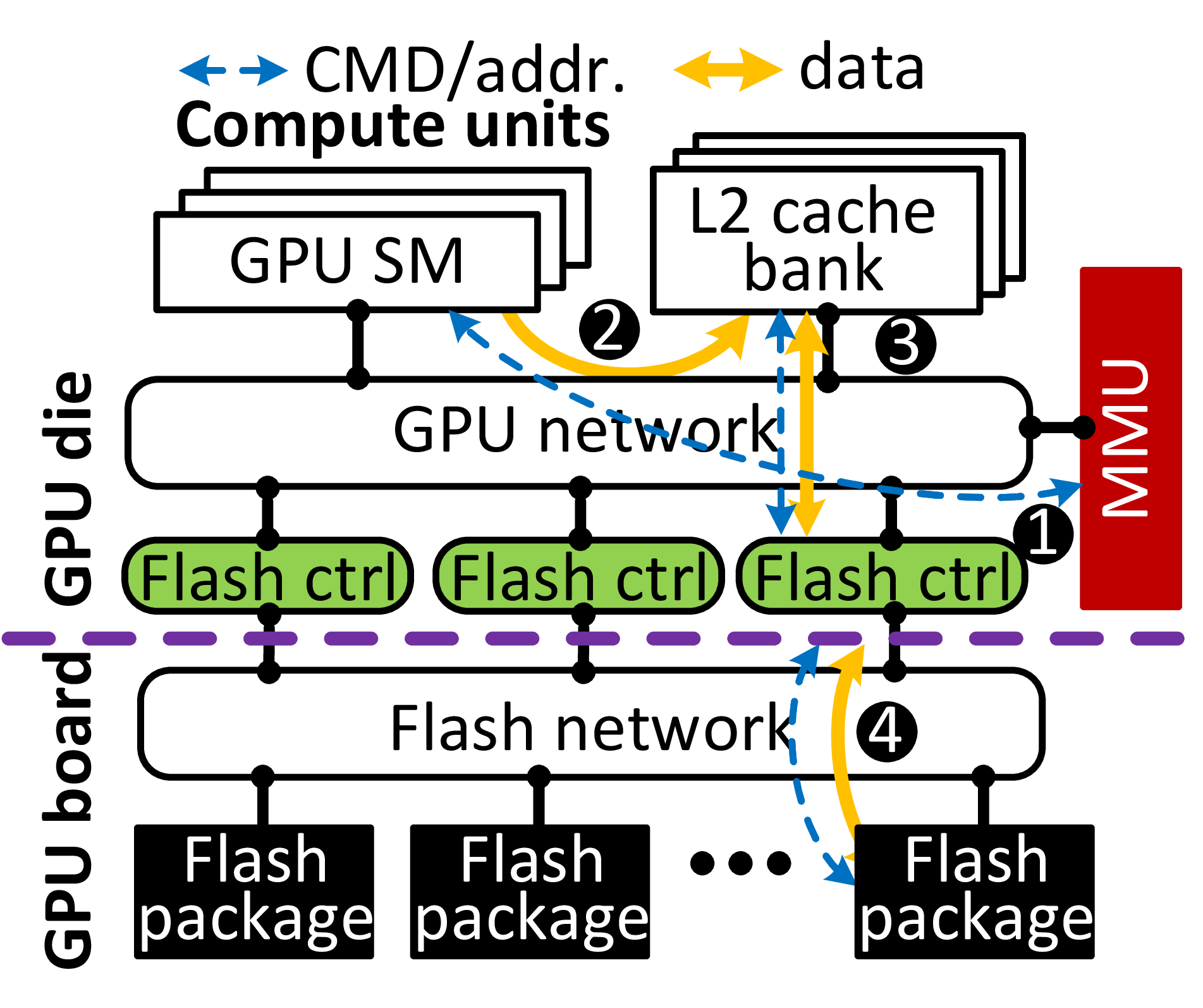}}}
\hspace{3pt}
\subfloat[Execution flow of memory accesses.]{\label{fig:exeflow}\rotatebox{0}{\includegraphics[width=0.71\linewidth]{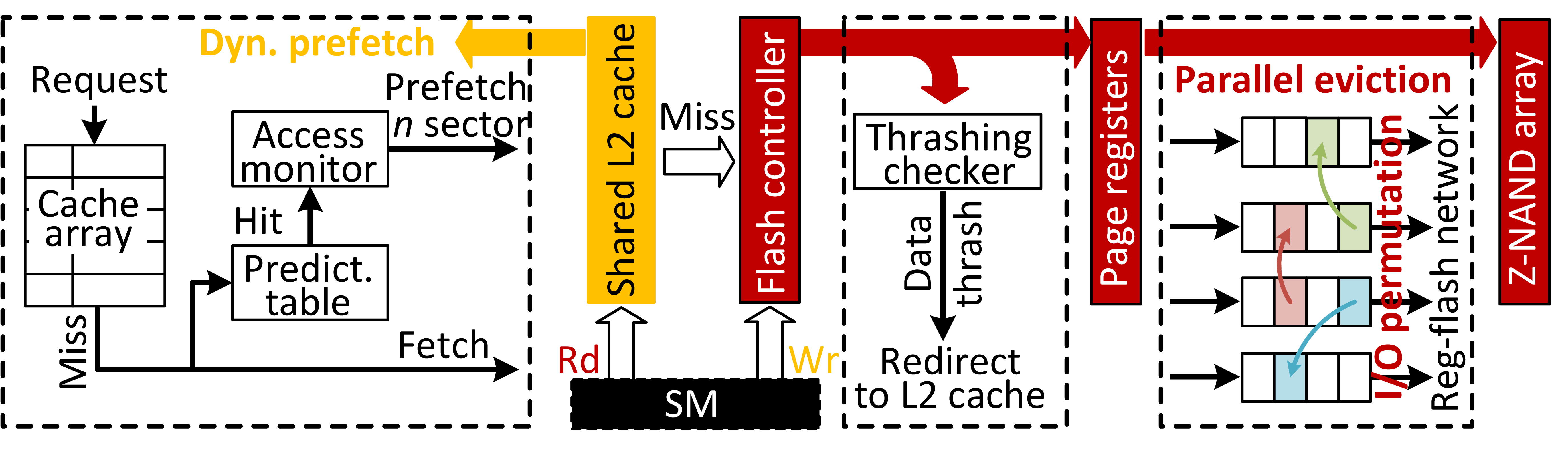}}}
\caption{\label{fig:overall}High-level view of ZnG and the execution flow of memory accesses.\vspace{-5pt}}
\vspace{-10pt}
\end{figure*}


\subsection{Key observations}
\label{sec:observation}
\noindent \textbf{FTL and GPU-SSD interconnection.}
Figure \ref{fig:motiv7-latency} analyzes multiple components that contribute the memory access latency and compares the latency breakdown between traditional GPU memory subsystem and HybridGPU. The SSD engine, which performs FTL, accounts for 67\% of the total memory access latencies. This is because the SSD engine in commercial SSDs has limited computing power to process the memory requests, which are simultaneously generated by the massive GPU cores. As a result, a large number of memory requests can be stalled by the SSD engine. In addition to the SSD engine issue, the network latency (between L2 cache and SSD engine) is significant, compared to the traditional GPU memory subsystem. This is because the SSD controller only employs 2$\sim$5 low-power embedded processor cores \cite{SSDcontroller}, which are unable to send/receive many memory requests in parallel. Thus, the SSD controller also becomes the performance bottleneck of serving memory requests. 

\noindent \textbf{Z-NAND access granularity.}
We perform a simulation-based study to analyze the impact of replacing all GPU on-board DRAMs with Z-NAND flash packages. For the sake of brevity, we assume there is no performance penalty introduced by the SSD controller. In this study, we configure the GTX580-like GPU model \cite{nvidia2012nvidia} by using a GPU simulation framework, MacSim \cite{kim2012macsim}. The important configuration details are listed in Table \ref{tab:sys-config}. Figure \ref{fig:motiv-baselinevsflashGPU} shows the performance degradation, imposed by direct Z-NAND flash accesses, under the execution of 12 real GPU workloads \cite{nai2015graphbig, che2009rodinia, pouchet2012polybench}. One can observe from this figure that the performance degradation can be as high as 28$\times$. This is because the memory access size in GPU is 128B, which is much smaller than the minimum access granularity of Z-NAND flash (4KB). Therefore, 97\% of flash access bandwidth is underutilized when serving the requests. 

\noindent \textbf{Workload characteristics.}
We evaluate the impact of direct Z-NAND accesses on large-scale data analysis applications. 
Figure \ref{fig:motiv-readreaccess} shows the number of memory requests that repeat accessing the same pages. Each Z-NAND page, on average, is repeatedly read by 42 times, across all the workloads. This observation indicates that it is still beneficial to buffer the re-accessed data for future fast accesses. \newedit{In addition to the read operations, we also collect the number of write requests that target to the same Z-NAND pages, called \textit{write redundancy} in this work. The results are shown in Figure \ref{fig:motiv-writeamp}.} Each Z-NAND page, on average, receive 65 times of write operations. Serving such write requests in flash pages can dramatically shorten the lifetime of Z-NAND. Thus, it is essential to have a buffer to accommodate write requests.

\subsection{The high-level view of ZnG}
\label{sec:highlevelview}
\noindent \textbf{Putting Z-NAND flash close to GPU.}
Figure \ref{fig:highlevelview} shows an architectural overview of the proposed ZnG. Compared to HybridGPU, ZnG removes the components of the request dispatcher, the SSD controller and the data buffer (which are placed between the GPU L2 cache and the Z-NAND flash). Instead, the underlying flash network is directly attached to the GPU interconnect network through the flash controllers. While the flash controllers manage the I/O transactions of the underlying Z-NAND, ZnG integrates a request dispatcher in each flash controller to interact with the GPU interconnect network in sending/receiving the request packets. There exist two root causes that ZnG does not directly attach Z-NAND packages to the GPU interconnect network. First, Z-NAND packages leverage Open NAND Flash Interface (ONFI) \cite{workgroup2011open} for the I/O communication whose frequency and hardware (electrical lane) configurations are different from those of the GPU interconnect network. Second, since a bandwidth capacity of the GPU interconnect network much exceeds total bandwidth brought by all the underlying Z-NAND packages, directly attaching Z-NAND packages to the GPU interconnect network can significantly underutilize the network resources. Thus, we employ a mesh structure as the flash network, which can meet the bandwidth requirement of Z-NAND packages by increasing the frequency and link widths, rather than leveraging the existing GPU interconnect network.

\noindent \textbf{Zero-overhead FTL.}
As the SSD controller is removed from ZnG, we offload FTL to other hardware components. GPU's MMU can be a good candidate component to implement the FTL as all memory requests leverage the MMU to translate their virtual addresses to memory logical addresses. We can achieve a zero-overhead FTL if the MMU directly translates the virtual address of each memory request to flash physical address. However, MMU does not have a sufficient space to accommodate all the mapping information of FTL. An alternative solution is to revise the internal row decoder of each Z-NAND plane to remap a request's address to a wordline of Z-NAND flash array, which is inspired by \cite{decoder}. While this approach can eliminate the FTL overhead, reading a page requires searching the row decoders of all Z-NAND planes, which introduces huge Z-NAND access overhead. To address these challenges, ZnG collaborates such two approaches. Our key observation is that a wide spectrum of the data analysis workloads is read-intensive, which generates only a few write requests to Z-NAND. Thus, we split FTL's mapping table into a read-only block mapping table and a log page mapping table. To reduce the mapping table size, the block mapping table records the mapping information of a flash block rather than a page. This design in turn reduces the mapping table size to 80 KB, which can be placed in the MMU. While read requests can leverage the read-only block mapping table to find out its flash physical address, this mapping table cannot remap incoming write requests to new Z-NAND pages. To address this, we implement a log page mapping table in the flash row decoder. The MMU can calculate the flash block address of the write requests based on the block mapping table. We then forward the write requests to the target flash block. The flash row decoder remaps the write requests to a new page location in the flash block. Once all the spaces of Z-NAND are used up, we allocate a GPU helper thread to reclaim flash block(s) by performing garbage collection, which is inspired by \cite{vijaykumar2016case}.

\begin{figure*}
\centering
\subfloat[Mapping tables.]{\label{fig:mapping_table}\rotatebox{0}{\includegraphics[width=0.36\linewidth]{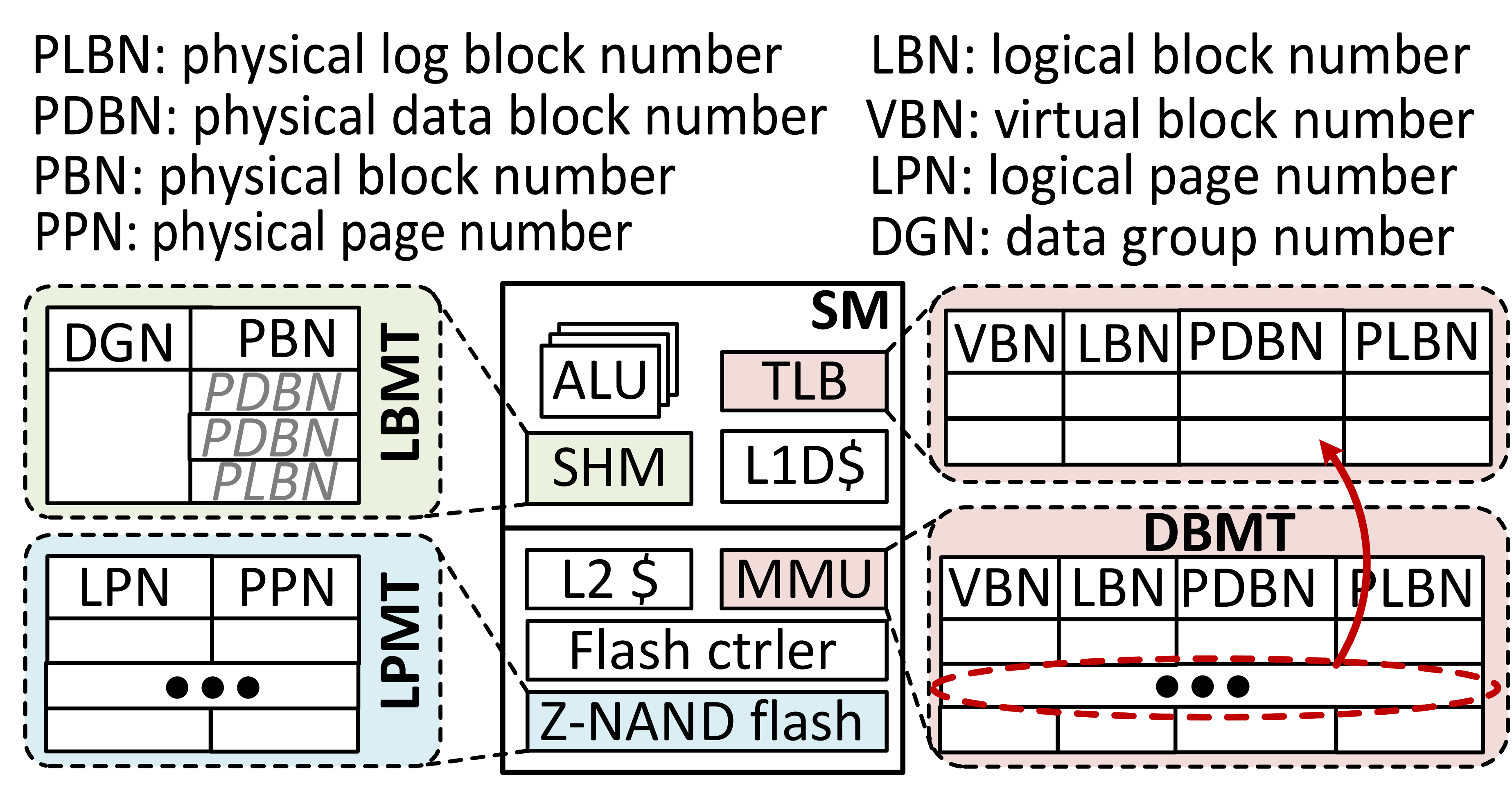}}}
\hspace{2pt}
\subfloat[Row decoder modification.]{\label{fig:row_dec}\rotatebox{0}{\includegraphics[width=0.61\linewidth]{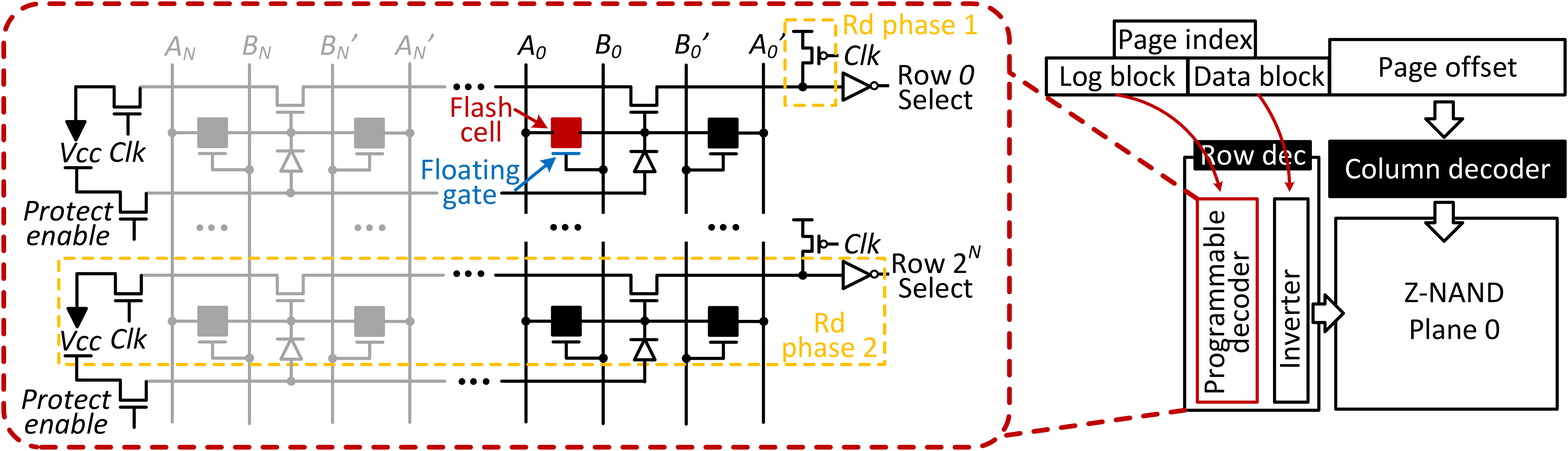}}}
\caption{\label{fig:ftl}Zero-overhead FTL design.\vspace{-5pt}}
\vspace{-15pt}
\end{figure*}

\noindent \textbf{Putting it all together.}
To make a memory request correctly access the corresponding Z-NAND, the memory requests, generated by GPU SMs, firstly leverage TLB/MMU to translate their logical addresses to flash physical addresses (\whitecircled{1}) (cf. Figure \ref{fig:highlevelview}). GPU L1 and L2 caches are indexed by the flash physical address. If a memory request misses in the L2 cache (\whitecircled{2}), the L2 cache sends the memory request to one of the flash controllers (\whitecircled{3}). The flash controller decodes the physical address to find the target flash plane and converts the memory requests into a sequence of flash commands (\whitecircled{4}). The target Z-NAND serves such memory request by using the row decoder to activate the wordline and sensing data from the flash array.

\subsection{The optimizations for high throughput}
As explained in Section \ref{sec:observation}, simply removing the internal DRAM buffer imposes huge performance degradation in a GPU-SSD system. To address the degradation, ZnG assigns GPU L2 cache and Z-NAND internal registers as read and write buffers, respectively, to accommodate the read and write requests. Figure \ref{fig:exeflow} shows the high-level view of our proposed design. Specifically, we increase the L2 cache capacity by replacing SRAM with non-volatile memory, in particular STT-MRAM to accommodate as many read requests as possible. While STT-MRAM can increase the L2 cache capacity by 4 times, its long write latency makes it infeasible to accommodate the write requests \cite{zhang2019fuse}. We then increase the number of registers in Z-NAND to shelter the write requests.

\noindent \textbf{Read prefetch.}
L2 cache can better serve the memory requests if it can accurately prefetch the target data blocks from the underlying Z-NAND. We propose a predictor to speculate spatial locality of the access pattern, generated by user applications. If the user applications access the continuous data blocks, the predictor informs the L2 cache to prefetch data blocks. As the limited size of L2 cache cannot accommodate all prefetched data blocks, we also propose an access monitor to dynamically adjust the data sizes in each prefetch operation.

\noindent \textbf{Construct fully-associative flash registers.} 
Z-NAND plane allows to employ a few flash registers (i.e., 8) due to space constraints. The limited number of flash registers may not be sufficient to accommodate all write requests based on workload execution behaviors. To address this, we propose to group the flash registers across all flash planes in the same Z-NAND flash package to serve as a fully-associative cache, such that memory requests can be placed in anywhere within the flash registers. However, this requires all the flash registers to connect to both I/O ports and all flash arrays in a Z-NAND package, which introduce high interconnection cost. To address this issue, we simplify the interconnection by connecting all the flash registers with a single bus, and only the buses are connected to the I/O ports and flash planes. We also propose a thrashing checker to monitor if there is cache thrashing in the limited flash registers. If so, ZnG pins a few L2 cache space to place the excessive dirty pages.

\section{Implementation}
\label{sec:implementation}
\begin{figure*}
\centering
\subfloat[The design of dynamic read prefetch module.]{\label{fig:prefetch}\rotatebox{0}{\includegraphics[width=0.4\linewidth]{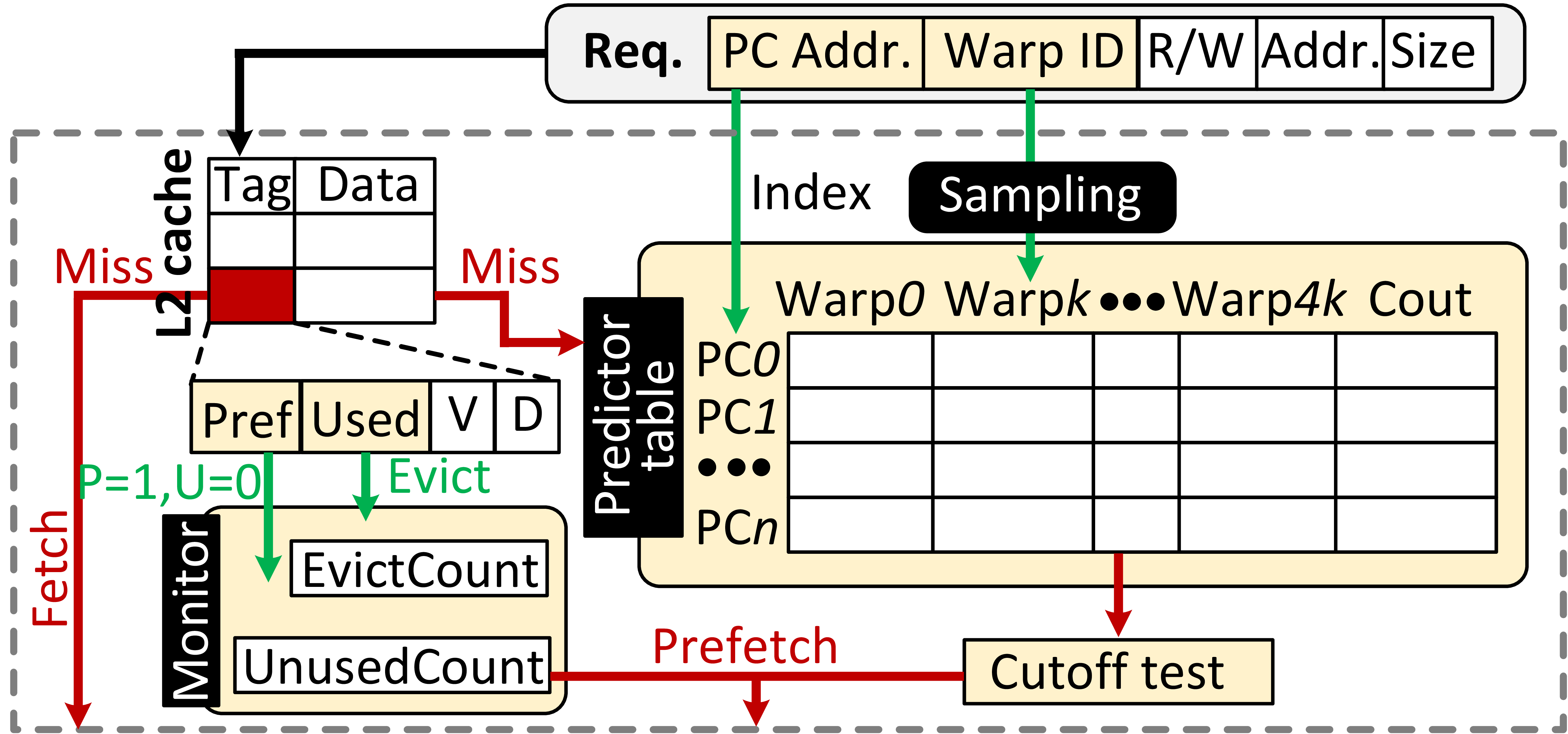}}}
\hspace{1pt}
\subfloat[Asymmetric Z-NAND writes.]{\label{fig:motiv4}\rotatebox{0}{\includegraphics[width=0.26\linewidth]{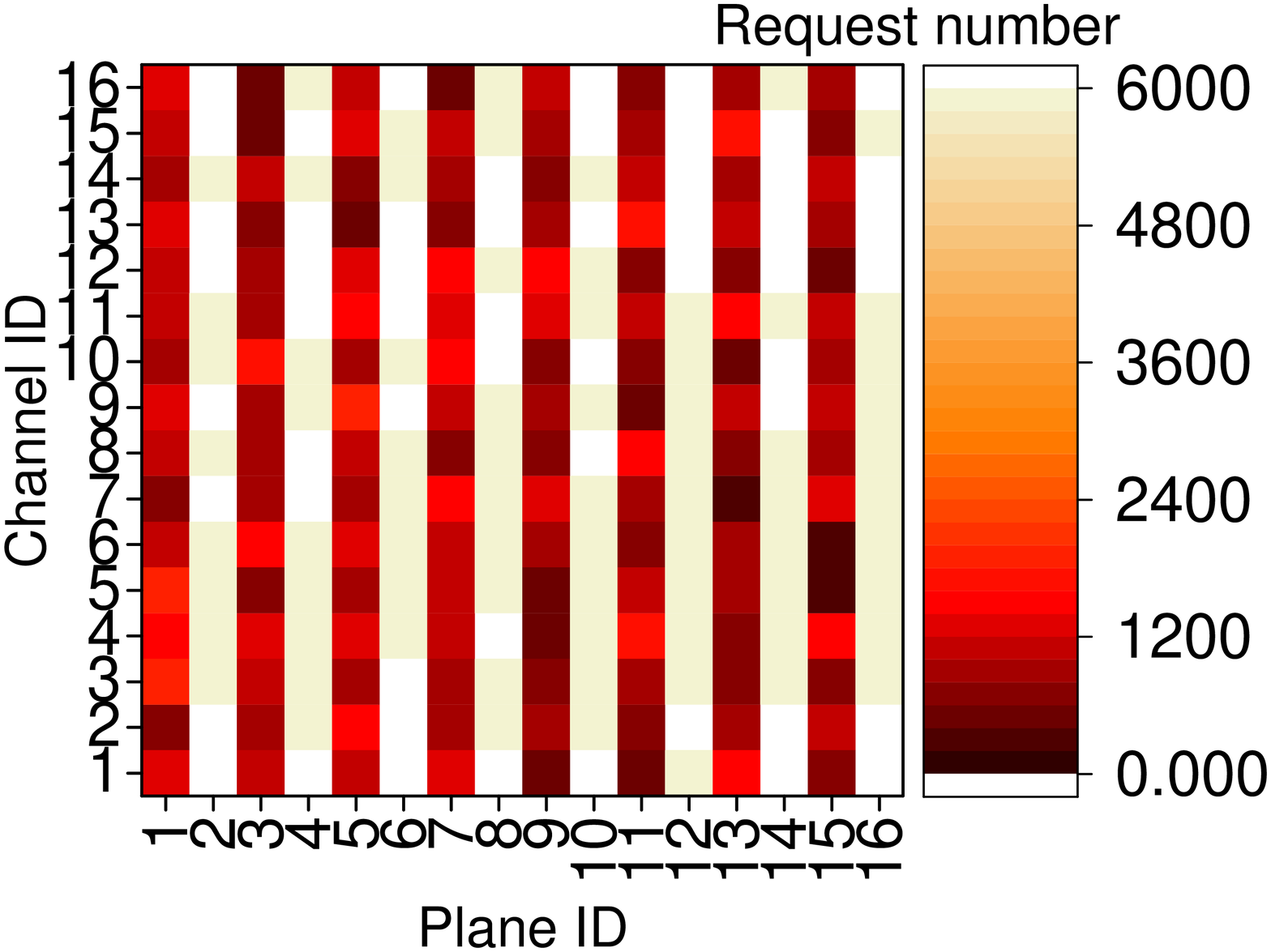}}}
\hspace{1pt}
\subfloat[Software I/O permutation.]{\label{fig:sw_permutation}\rotatebox{0}{\includegraphics[width=0.3\linewidth]{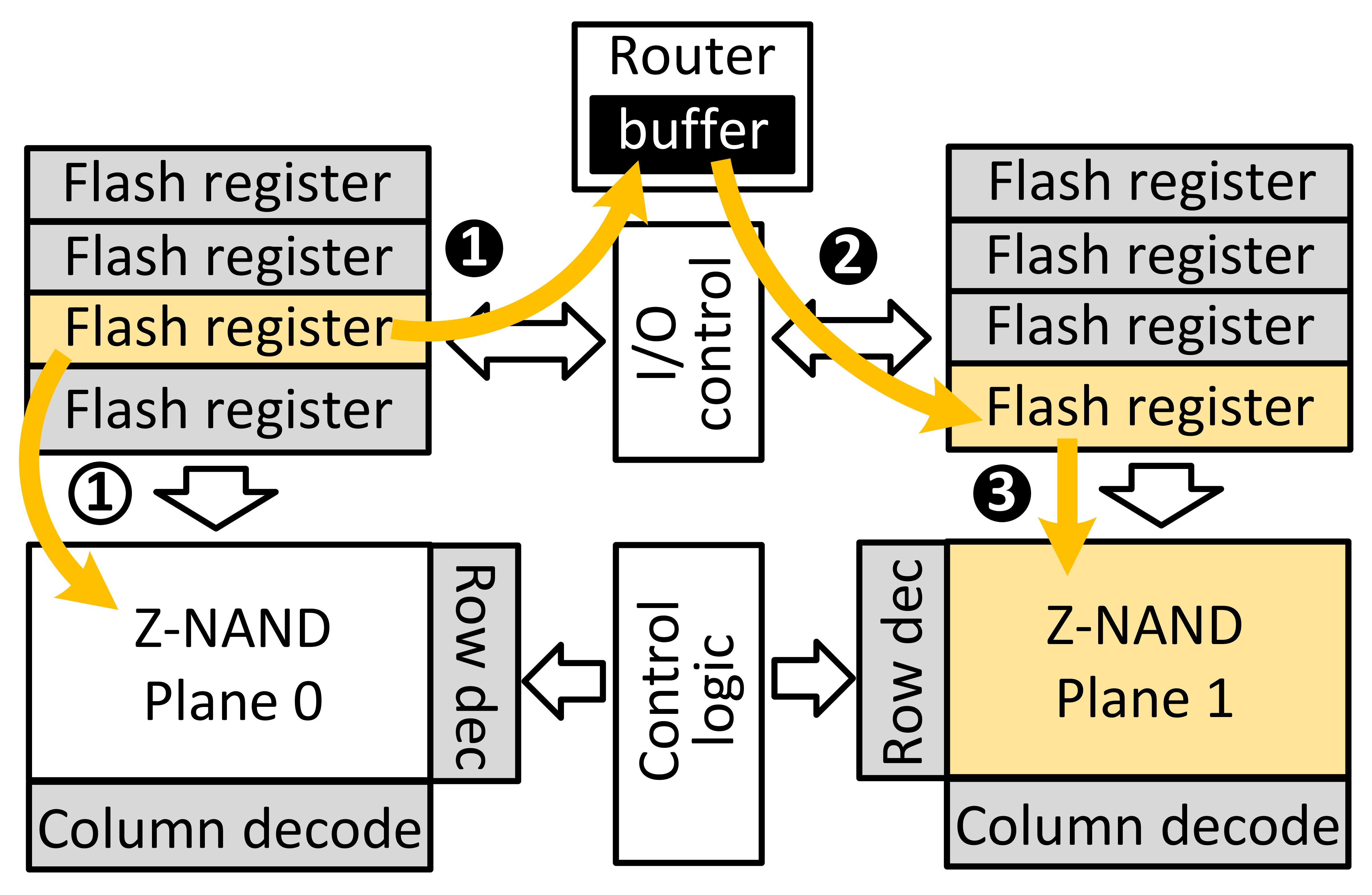}}}
\caption{\label{fig:rdwr_opt}Dynamic read prefetch design and S/W I/O permutation design.\vspace{-5pt}}
\vspace{-15pt}
\end{figure*}

\subsection{Zero-overhead FTL design}
\noindent \textbf{Overall implementation of zero-overhead FTL.}
Figure \ref{fig:mapping_table} shows an overview of the FTL implementation across GPU's MMU/TLB, the shared memory, and the flash row decoder. Our MMU implementation adopts a two-level page table based on a real GPU MMU implementation \cite{armgpummu}. The page table works as a data block mapping table (DBMT), whose entries store virtual block number (VBN), logical block number (LBN), physical log block number (PLBN) and physical data block number (PDBN). Among these, VBN is the data block address of the user applications in the virtual address space. LBN is a global memory address, while PDBN and PLBN are Z-NAND's flash addresses. The physical data block sequentially stores the read-only flash pages. If memory requests access read-only data, the requests can locate the positions of the target data from PDBN by referring their virtual addresses as an index. On the other hand, the write requests are served by physical log blocks. We employ a logical page mapping table (LPMT) for each physical log block to record such information. If memory requests access a modified data block, they need to refer to LPMT to find out the physical location of the target data. Each LPMT is stored in the row decoder of the corresponding log block. We will explain the detailed implementation of LPMT shortly. The TLB is employed to accelerate the flash address translation. It buffers the entries of DBMT, which are frequently inquired by the GPU kernels. Note that the physical log blocks come from SSD's over-provisioned space and therefore, the blocks are not accounted in the address space of Z-NAND \cite{park2008reconfigurable}. 
Considering the limited SSD's over-provisioned space, we group multiple physical data blocks to share a physical log block. We also create a log block mapping table (LBMT) (in the shared memory) to record such mapping information (cf. Figure \ref{fig:mapping_table}). 

While MMU can perform the address translation, it does not support other essential functionalities of FTL such as the wear-levelling algorithm and garbage collection. We further implement these functions in a GPU helper thread. Specifically, when all the flash pages in a physical log block have been used up, the GPU helper thread performs garbage collection, which merges the pages of all physical data blocks and the physical log blocks. It then selects empty physical data blocks based on wear-levelling algorithm to store the merged pages. The GPU helper thread lastly updates the corresponding information in the LBMT and the DBMT. 

\noindent \textbf{Overall implementation of LPMT in row decoders.}
Figure \ref{fig:row_dec} shows the implementation details of a programmable row decoder in Z-NAND. The flash controller converts each memory request to the corresponding physical log block number, physical data block number and page index, and sends them to the row decoder. To serve a read request, the programmable decoder looks up LPMT for the target data. If the target data hits in LPMT, the programmable decoder activates the corresponding wordline based on the page mapping information of LPMT. Otherwise, the row decoder activates the wordline based on the page index and the data block number. On the other hand, to serve a write request, a write operation is performed by selecting a free page in the physical log block to program the new data, and the new mapping information is recorded in LPMT. As an in-order programming is only allowed in Z-NAND, we can use a register to track the next available free page number in the physical log block.

The programmable decoder contains wordlines as many as those of Z-NAND flash array. Each wordline of the programmable decoder connects to \textit{2N} flash cells and \textit{4N} bitlines ($A_{0}$$\sim$$A_{N}$, $B_{0}$$\sim$$B_{N}$, $A_{0}^{'}$$\sim$$A_{N}^{'}$, $B_{0}^{'}$$\sim$$B_{N}^{'}$), where N is the physical address length.  
The page mapping information of the LBPT is programmed in the flash cells of the programming decoder by activating the corresponding wordlines and bitlines. The detailed steps of a write operation in programmable decoder are as follows: 1) the programmable decoder activates its wordline corresponding to the free page; 2) the values of page index, such as ``1'' and ``0'', are converted as a high and low voltage, respectively. Such voltage is applied to $B_{0}$$\sim$$B_{N}$, while the inverse voltage is applied to $B_{0}^{'}$$\sim$$B_{N}^{'}$. Thus, the voltages on $B$ and $B^{'}$ program the flash cells; 3) for other rows, the ``protect'' gates are enabled to drive high voltages to the drain selectors, which can avoid program disturbance. 
The programmable decoder operates as a content addressable memory (CAM) to search if the target data exists in a physical log block. Specifically, the search procedure includes two phases, which are shown in Figure \ref{fig:row_dec}. In phase 1, the clock signal disables the gates, and the wordlines are charged with a high voltage. In phase 2, the voltages converted from page index are applied to $A_{0}$$\sim$$A_{N}$ and $A_{0}^{'}$$\sim$$A_{N}^{'}$. If the page index matches with the values, stored in any row, the gates between A and $A^{'}$ are enabled to dicharge the corresponding wordline of the programmable decoder (which activates a row of the flash array). Otherwise, the wordlines keep the high voltage to disable the row selection.  
   
\begin{figure*}
\centering
\subfloat[Simple H/W permutation.]{\label{fig:hw_permutation1}\rotatebox{0}{\includegraphics[width=0.24\linewidth]{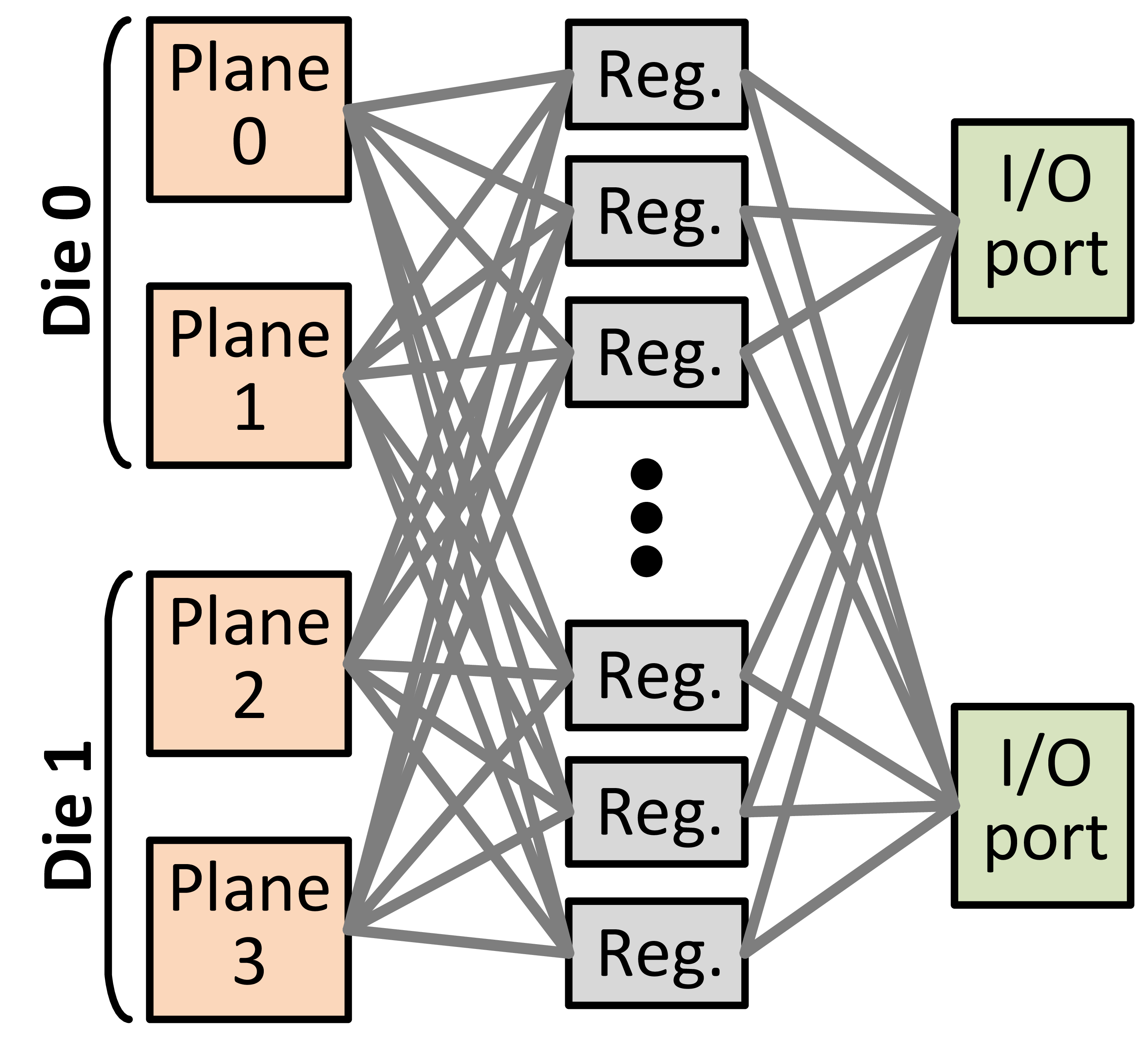}}}
\hspace{2pt}
\subfloat[H/W permutation with low hardware cost.]{\label{fig:hw_permutation2}\rotatebox{0}{\includegraphics[width=0.72\linewidth]{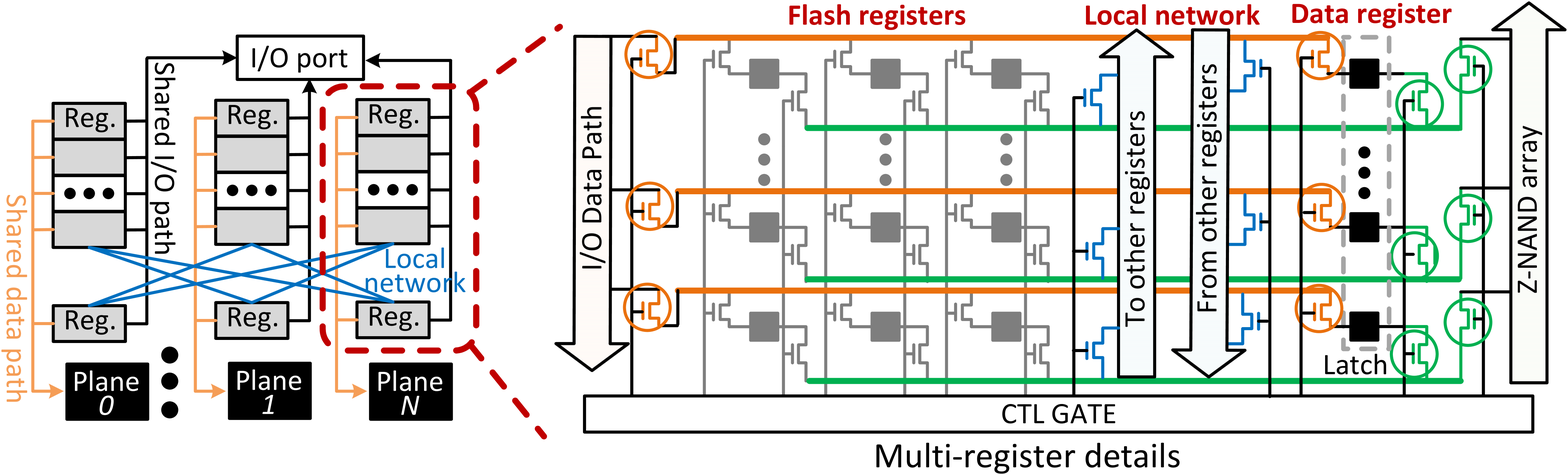}}}
\caption{\label{fig:ftl}FTL/SSD engine design.\vspace{-5pt}}
\vspace{-15pt}
\end{figure*}

\subsection{Read optimization}
\label{sec:rdopt}
Figure \ref{fig:prefetch} shows an overview of the proposed dynamic read prefetch to improve L2 cache utilization. Our design includes three components: 1) a predictor to speculate data locality; 2) an extension of L2 cache's tag array to track the status of data accesses; and 3) our access monitor to dynamically adjust the granularity of data prefetch.

\noindent \textbf{Predictor table.}
We design the predictor in L2 cache to record the access history of the read requests and speculate the memory access pattern based on the program counter (PC) address of each thread, which is inspired by \cite{zhang2019flashgpu}. The key insight behind this design is that the memory requests, generated from the LD/ST instructions of the same PC address, exhibits the same access patterns. The implementation of our predictor requires extending the memory requests with an extra field to record the PC address and warp ID. Our predictor also contains a predictor table, whose entries are indexed by the PC addresses. \newedit{We allocate 512 entries in the predictor table as default, which is the same to \cite{zhang2019flashgpu}. Each entry of the predictor table contains a few fields for different warps to store the logical page number that they are accessing. Fundamentally, we track the accesses of five representative warps. Each entry also includes a 4-bit counter to store the number of re-accesses to the recorded pages.} For example, if the warp 0 generates a memory request based on PC address 0 and the request targets to the same page as what is recorded in the predictor table, the counter increases by one. Otherwise, if the request accesses a page different from the page number (recorded in the predictor table), the counter decreases by one, and the new page number is filled in the corresponding field of the predictor table. When there is a cache miss, a cutoff test of read prefetch checks the predictor table by referring to the PC address of the memory request. If the counter value is higher than a threshold (i.e., 12), we perform a read prefetch.

\noindent \textbf{Extension of L2 cache's tag array.}
We extend each entry of L2 cache's tag array with the fields of an accessed bit and a prefetch bit. These two fields are used to check if the prefetched data were early evicted due to the limited L2 cache space. Specifically, we use the prefetch bit to identify whether the data stored in the cache line is filled by a prefetch, while the accessed bit records if a cache line has been accessed by a warp. When a cache line is evicted, the prefetch and accessed bits are checked. If the cache line is filled by a prefetch but has not been accessed by a warp, this indicates a read prefetch may introduce L2 cache thrashing.  

\noindent \textbf{Access monitor.}
To avoid early eviction of the prefetched data and improve the utilization of L2 cache, we propose an access monitor to dynamically adjust the access granularity of data prefetch. If a cache line is evicted, our access monitor updates its evict counter and unused counter by referring to aforementioned prefetch and accessed bits. We calculate a waste ratio of data prefetch by dividing the unused counter with the evict counter. \newedit{If the waste ratio is higher than a high threshold, we decrease the access granularity of data prefetch by half. If the waste ratio is lower than a low threshold, we increase the access granularity by 1KB. To determine the optimal threshold configurations, we performed an evaluation by sweeping different values of the high and low thresholds (cf. Section \ref{sec:sensitive-test}), and observed that ZnG can achieve the best performance by configuring the high and low thresholds as 0.3 and 0.05, respectively. Thus, we adopt these values by default.}

\subsection{Write optimization}
We analyze write patterns in different Z-NAND planes by executing a multi-application workload \texttt{betw-back} \cite{nai2015graphbig, che2009rodinia} and depict the results in Figure \ref{fig:motiv4}. One can observe from the figure that different Z-NAND planes from different channels experience different number of writes.  There are two root causes behind these write patterns: 1) SSD controller redirects the requests of different applications to access different flash planes, which can help reduce write amplification \cite{kang2014multi}; 2) an application may exhibit asymmetric accesses to different pages. Due to asymmetric writes on Z-NAND planes, a few flash registers can stay in idle, while other flash registers suffer from a data thrashing issue. To address this challenge, we may group different flash registers together to serve the write requests, such that data can be placed in anywhere of the flash registers. This register grouping can make the flash registers fully utilized. We propose simple software and hardware solutions, called \textit{SWnet} and \textit{FCnet}, respectively, which are shown in Figures \ref{fig:sw_permutation} and \ref{fig:hw_permutation1}. As shown in Figure \ref{fig:sw_permutation}, the flash controller in SWnet can directly control a flash register to write its data to the local flash plane 0 (\blkcircled{1}). If the flash register needs to write its data to the remote flash plane 1, the flash controller leverages a router in the flash network to copy the data to its internal buffer (\whitecircled{1}) and redirect the data to a remote flash register (\whitecircled{2}). Once data is available in the remote flash register, the flash controller coordinates the remote flash register to write such data to the flash plane 1 (\whitecircled{3}). Note that SWnet does not require any hardware modification on existing flash architectures. However, it needs data migration between the two flash registers and consumes flash network bandwidth. 

On the other hand, our simple hardware solution, FCnet, builds a fully-connected network to make all flash registers directly connect to the I/O ports and Z-NAND planes (cf. Figure \ref{fig:hw_permutation1}). While this fully-connected network can maximize the internal parallelism within a Z-NAND package, it needs a large number of point-to-point wire connections.
We optimize the hardware solution by connecting the flash registers to I/O ports and Z-NAND planes with a hybrid network. The new solution, called \emph{Network-in-Flash} (NiF), can reduce hardware cost and achieve high performance. Figure \ref{fig:hw_permutation2} shows our implementation details of NiF. Our key insight is that either an I/O port or a Z-NAND plane can only serve the data from a single flash register. Thus, in NiF, all flash registers from the same flash plane are connected to two buses. A bus is extended to connect the I/O port as \emph{a shared I/O path}, while another bus is connected to its local Z-NAND plane as \emph{a shared data path}. The control logic can select a flash register to use the I/O path by turning on the switch gate, while it can simultaneously select another flash register to access the local Z-NAND plane. In this design, a flash register cannot directly access a remote flash plane. Instead, we assign one flash register from the group of flash registers, referred to as data register, in connecting to other groups of flash registers via \emph{a local network}. If a flash register needs to write its data to a remote Z-NAND plane, such flash register firstly moves the data to the remote data register, and then the remote data register evicts the data to the remote Z-NAND plane. The bus structure is aimed to save costs of building a network inside a flash package. Although it still needs to migrate data between two registers, the data migration in NiF does not occupy the flash network. It also allows migrating multiple data in local network simultaneously, which can achieve a better degree of internal parallelism than SWnet.

\section{Evaluation}
\label{sec:evaluation}
\subsection{Experiment}
\noindent \textbf{Simulation methodology.}
We use SimpleSSD \cite{jung2017simplessd} and MacSim \cite{kim2012macsim} to model an SSD and GPU, similar to a 800GB ZSSD \cite{koh2018exploring} and NVIDIA GTX580 \cite{nvidia2012nvidia}, respectively; Table \ref{tab:sys-config} explains the details of each configuration. Besides, we increase the L2 cache size to match with the configuration of the state-of-the-art GPU, NVIDIA GV100 \cite{gv100}. STT-MRAM that we simulate can increase the capacity of L2 cache by 4$\times$, but its write latency is 5$\times$ than SRAM read latency \cite{zhang2019fuse}. We also configure the network widths of both HW-NiF and the flash network to 8B, which is 8$\times$ higher than traditional flash channel (1B). Lastly, we derive the latency model of Optane DC PMM from the evaluation of real devices \cite{izraelevitz2019basic}.

\begin{table}[]
\resizebox{\linewidth}{!}{
\begin{tabular}{|l|c|l|c|l|c|}
\hline
\multicolumn{2}{|c|}{\textbf{GPU}} & \multicolumn{2}{c|}{\textbf{Z-NAND array}} & \multicolumn{2}{c|}{\textbf{STT-MRAM L2 cache}} \\ \hline
\textbf{SM/freq.} & 16/1.2 GHz & \textbf{channel/package} & 16/1 & \textbf{size} & 24MB, shared \\ \hline
\textbf{max warps} & 80 per core & \textbf{die/plane} & 8/8 & \textbf{latency} & R:1-cycle, W:5-cycle \\ \hline
\multirow{3}{*}{\textbf{L1 cache}} & 1-cycle, 64-set & \textbf{block/page} & 1024/384 & \multicolumn{2}{c|}{\textbf{Flash network}} \\ \cline{2-6} 
 & 6-way, 48KB & \textbf{interface} & 800MT/s & \textbf{type} & mesh \\ \cline{2-6} 
 & LRU, private & \textbf{cell type} & SLC & \textbf{bus width} & 8B \\ \hline
\multirow{3}{*}{\textbf{L2 cache}} & 1-cycle, 6 banks & \textbf{register} & 2/8 per plane & \multicolumn{2}{c|}{\textbf{Optane DC PMM}} \\ \cline{2-6} 
 & 1024-set, 8-way & \textbf{I/O ports} & 2 per package & \textbf{tRCD/tCL} & 190/8.9ns \\ \cline{2-6} 
 & \newedit{LRU}, 6MB, shared & \textbf{HW-NiF width} & 8B & \textbf{tRP} & 763ns \\ \hline
\end{tabular}}
\caption{System configurations of ZnG. \label{tab:sys-config} \vspace{-10pt}}
\end{table}

\begin{table}[]
\resizebox{\linewidth}{!}{
\begin{tabular}{|l|l|c|c|l|l|c|c|}
\hline
Workload & \multicolumn{1}{c|}{\textbf{Suites}} & \textbf{read ratio} & \textbf{Kernel} & \multicolumn{1}{c|}{\textbf{Workload}} & \multicolumn{1}{c|}{\textbf{Suites}} & \textbf{read ratio} & \textbf{Kernel} \\ \hline
\textbf{betw} & \cite{nai2015graphbig} & 0.98 & 11 & \textbf{gc2} & \cite{nai2015graphbig} & 0.99 & 10 \\ \hline
\textbf{bfs1} & \cite{nai2015graphbig} & 0.95 & 7 & \textbf{sssp3} & \cite{nai2015graphbig} & 0.98 & 8 \\ \hline
\textbf{bfs2} & \cite{nai2015graphbig} & 0.99 & 9 & \textbf{deg} & \cite{nai2015graphbig} & 1 & 1 \\ \hline
\textbf{bfs3} & \cite{nai2015graphbig} & 0.88 & 10 & \textbf{pr} & \cite{nai2015graphbig} & 0.99 & 53 \\ \hline
\textbf{bfs4} & \cite{nai2015graphbig} & 0.97 & 12 & \textbf{back} & \cite{che2009rodinia} & 0.57 & 1 \\ \hline
\textbf{bfs5} & \cite{nai2015graphbig} & 0.99 & 6 & \textbf{gaus} & \cite{che2009rodinia} & 0.66 & 3 \\ \hline
\textbf{bfs6} & \cite{nai2015graphbig} & 0.97 & 7 & \textbf{FDT} & \cite{pouchet2012polybench} & 0.73 & 1 \\ \hline
\textbf{gc1} & \cite{nai2015graphbig} & 0.98 & 8 & \textbf{gram} & \cite{pouchet2012polybench} & 0.75 & 3 \\ \hline
\end{tabular}}
\caption{GPU benchmarks. \label{tab:workloads} \vspace{-5pt}}
\vspace{-15pt}
\end{table}

\noindent \textbf{GPU-SSD platforms.}
We implement seven different GPU-SSD platforms: \newedit{(1) \texttt{Hetero}: GPU and SSD are discrete devices attached to the host;} (2) \texttt{hybridGPU} \cite{zhang2019flashgpu}; (3) \texttt{Optane}: integrating Optane DC PMM in GPU by employing six memory controllers to connect six different Optane DC PMM; (4) \texttt{ZnG-base}: the baseline architecture of ZnG, which integrates the implementation of Section \ref{sec:highlevelview} without the read and write optimizations; (5) \texttt{ZnG-rdopt}: integrating the design of L2 cache in \texttt{ZnG-base}; \newedit{(6) \texttt{ZnG-wropt}: integrating the design of flash registers in \texttt{ZnG-base}; (7) \texttt{ZnG}: putting both \texttt{ZnG-rdopt} and \texttt{ZnG-wropt} into \texttt{ZnG-base}.}

\noindent \textbf{Workloads.}
We select a large number of applications from a graph analysis benchmark \cite{nai2015graphbig} and scientific benchmarks \cite{che2009rodinia, pouchet2012polybench}. The details of our evaluated workloads are provided in Table \ref{tab:workloads}. We then generate multi-app workloads by co-running a read-intensive workload and a write-intensive workload. Concurrently executing multiple applications can stress the memory subsystem in the GPU, while the complex access patterns generated by multiple applications can examine a robustness of the proposed techniques.

\begin{figure}
	\centering
	\vspace{-15pt}
	\includegraphics[width=1\linewidth]{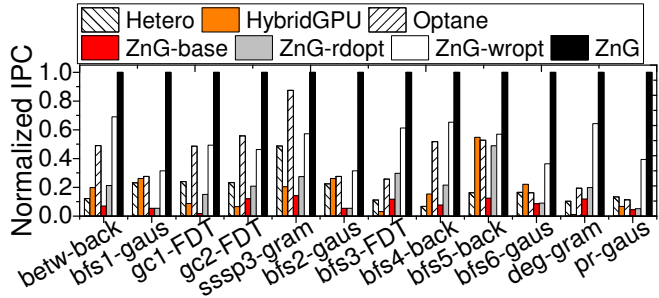}
	\vspace{-10pt}
	\caption{\label{fig:norm_ipc}Performance of different GPU architectures.\vspace{-5pt}}
\end{figure}

\begin{figure}
	\centering
	\includegraphics[width=1\linewidth]{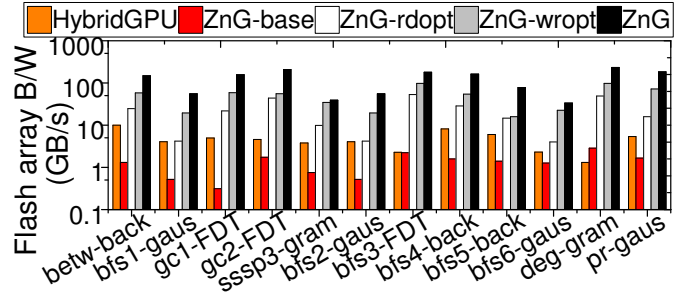}
	\vspace{-10pt}
	\caption{\label{fig:SSD_bw}Bandwidth analysis of flash arrays.\vspace{-5pt}}
	\vspace{-15pt}
\end{figure}

\subsection{Performance}
\noindent \textbf{IPC.}
Figure \ref{fig:norm_ipc} shows IPC values of different GPU-SSD platforms under various workload executions, and the values are normalized to \texttt{ZnG}'s IPCs. \newedit{Although the GPU in \texttt{Hetero} can enjoy high bandwidth of its internal GDDR5 DRAM, data initially resides in the external SSD, which takes a long delay to move from the SSD to the GPU. As shown in the figure, \texttt{HybridGPU} outperforms \texttt{Hetero} by 31\%, on average, under the workloads, \texttt{betw-back}, \texttt{bfs1-gaus}, \texttt{bfs2-gaus} and \texttt{bfs6-gaus}.} 
\texttt{Optane} improves the performance by 186\%, compared to \texttt{HybridGPU}, in overall. This is because \texttt{Optane} directly replaces DRAM with Optane DC PMM, whose accumulated bandwidth can be upto 39GB/s, while bandwidth of \texttt{HybridGPU} is limited by its SSD controller and internal memory bandwidth, which is only 5GB/s. As directly serving read/write requests with Z-NAND can drastically waste network bandwidth and flash bandwidth, both \texttt{ZnG-base} and \texttt{ZnG-rdopt} cannot catch up the performance of \texttt{HybridGPU}. \newedit{By merging all incoming small write requests in flash registers of Z-NAND, \texttt{ZnG-wropt} can effectively reduce the number of flash programming operations, which can improve its performance by 2.6$\times$, compared to \texttt{ZnG-rdopt}.} Lastly, by effectively buffering data in L2 cache and flash registers, \texttt{ZnG} can fully utilize its accumulated bandwidth, which is much higher than \texttt{Optane}; \texttt{ZnG} can achieve 1.9$\times$ higher bandwidth than \texttt{Optane}, on average. 

\noindent \textbf{Bandwidth of Z-NAND flash arrays.}
Figure \ref{fig:SSD_bw} shows the bandwidth of Z-NAND flash arrays measured from different GPU-SSD platforms. The average bandwidth of \texttt{HybridGPU} is only 4.2 GB/s due to constraints of limited flash network bandwidth and bandwidth of the internal DRAM buffer. Although \texttt{ZnG-base} employs a high-performance flash network, only a small amount of its flash array bandwidth is used for the incoming memory requests because of its large access granularity. As L2 cache can buffer a whole flash page and its accumulated bandwidth is much higher than the internal DRAM buffer in \texttt{HybridGPU}, \texttt{ZnG-rdopt} can improve the flash array bandwidth by 2.9$\times$, compared to \texttt{HybridGPU}. \newedit{However, \texttt{ZnG-rdopt} can be blocked by a few small write requests, as Z-NAND's write latency is 33$\times$ longer than its read latency. By buffering the write requests in the flash registers, the flash array bandwidth  of \texttt{ZnG-wropt} exceeds that of \texttt{ZnG-rdopt} by 137\%, on average. Lastly, \texttt{ZnG} can buffer the small write requests in both the L2 cache and flash registers. Therefore, it further increases the flash array bandwidth by 167\%, compared to \texttt{ZnG-wropt}, on average.}

\begin{figure}
	\centering
	\includegraphics[width=1\linewidth]{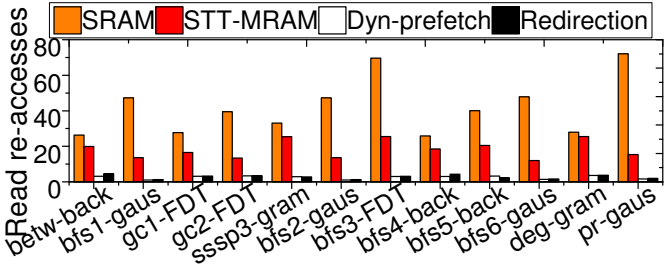}
	\vspace{-10pt}
	\caption{\label{fig:read-reaccess}The number of read re-accesses in flash arrays.\vspace{-5pt}}
	\vspace{-10pt}
\end{figure}

\begin{figure}
	\centering
	\includegraphics[width=1\linewidth]{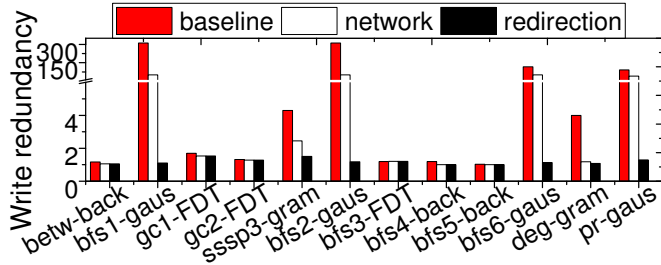}
	\vspace{-10pt}
	\caption{\label{fig:write-amp}Write redundancy in flash arrays.\vspace{-5pt}}
	\vspace{-15pt}
\end{figure}

\begin{figure}
	\centering
	\vspace{-10pt}
	\includegraphics[width=1\linewidth]{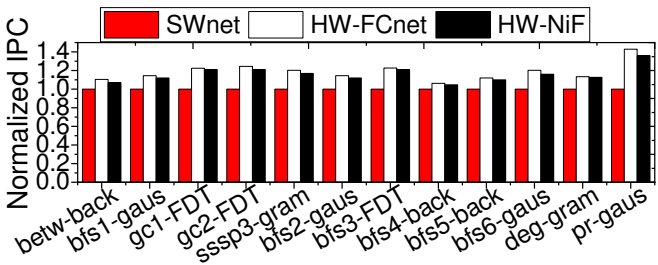}
	\vspace{-10pt}
	\caption{\label{fig:norm_ipc_nif}Performance analysis of flash register designs.\vspace{-5pt}}
\end{figure}

\subsection{Effectiveness of Optimizations}
\noindent \textbf{Read re-accesses.}
Figure \ref{fig:read-reaccess} demonstrates how our proposed technique can reduce the number of read re-accesses in the Z-NAND flash array. As STT-MRAM can increase L2 cache capacity to accommodate more read requests, replacing SRAM with STT-MRAM in L2 cache can reduce the average number of read re-accesses by 55\%. Employing our proposed dynamic prefetching technique in L2 cache can further reduce the number of read re-accesses by 87\%. This is because the dynamic prefetching technique can better utilize the limited space of L2 cache by only allocating the cache space for data that will be accessed soon. Even though pinning L2 cache space for the excessive write requests (\texttt{Redirection}) can reduce available L2 cache size to accommodate the read requests, we observe that the number of read re-accesses only increases by 11\%, compared to \texttt{Dyn-prefetch}. 

\noindent \textbf{Write redundancy.}
Figure \ref{fig:write-amp} shows how our proposed flash register design can effectively reduce the write redundancy in Z-NAND. The average number of write redundancy in \texttt{baseline} is 51 as \texttt{baseline} employs 8 flash registers to buffer and merge the massive small write requests. By employing NiF (\texttt{network}), the flash registers from different planes in the same flash package can be grouped together to serve incoming write requests. By improving the utilization of flash registers, \texttt{network} can reduce such write redundancy by 46\%. However, write-intensive scientific workloads such as \texttt{gaus} can generate excessive write requests to the limited flash registers, which makes the flash register thrashing. By redirecting the write requests to the pinned L2 cache space, we can mitigate the negative impact of thrashing issue, which reduces the write redundancy to 1.2, on average.

\noindent \textbf{Network in flash.}
Figure \ref{fig:norm_ipc_nif} compares the performance of using different network designs in Z-NAND flash packages. Employing hardware-based fully-connected network (\texttt{HW-FCnet}) can achieve 19\% higher performance than the software-based solution (\texttt{SWnet}). However, the building cost of a fully-connected network in the flash package is not affordable. Our NiF solution (\texttt{HW-NiF}) can achieve 98\% of the performance in \texttt{HW-FCnet}, while it can reduce the cost of network construction in Z-NAND flash packages.


\begin{figure}
\centering
\subfloat[Performance of multi-applications.]{\label{fig:multiapp1}\rotatebox{0}{\includegraphics[width=0.69\linewidth]{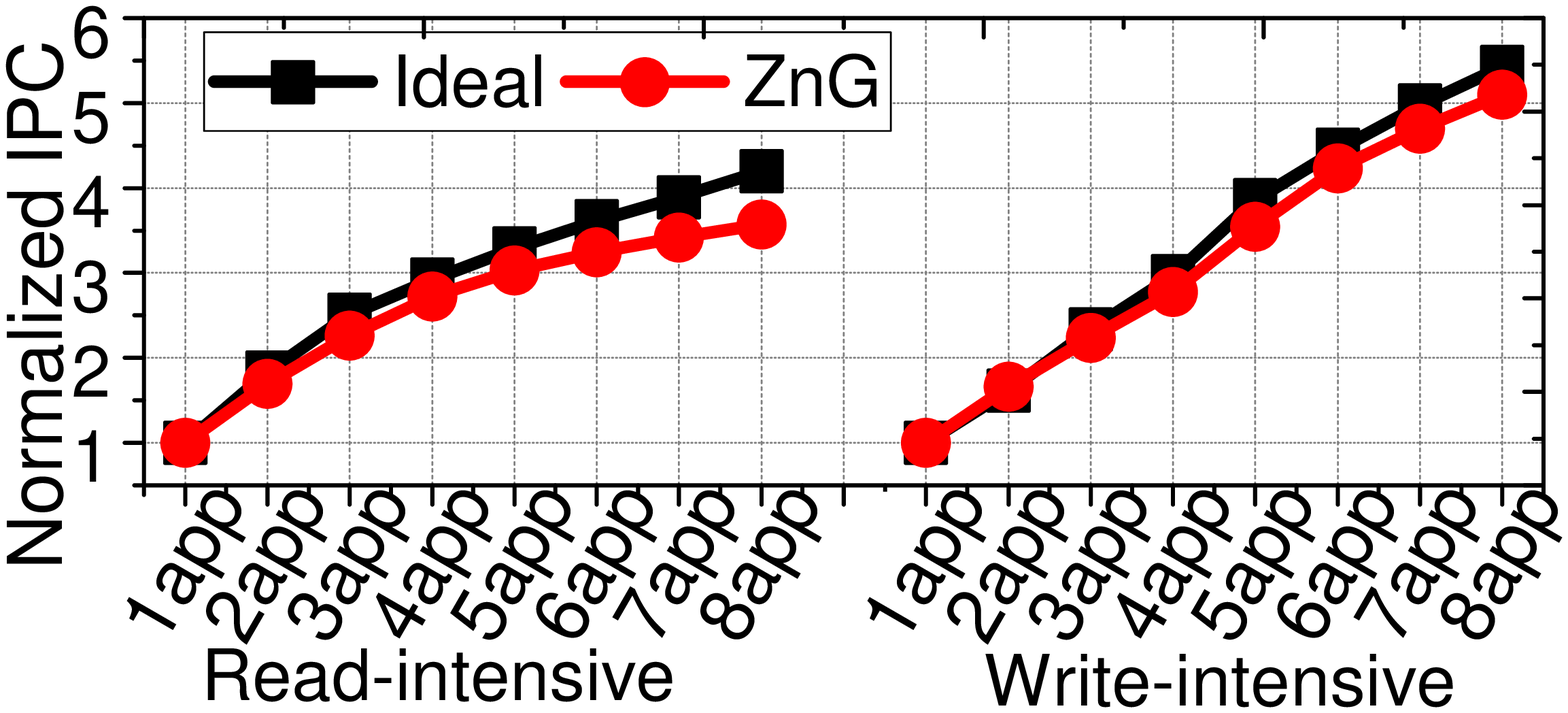}}}
\hspace{2pt}
\subfloat[Accuracy.]{\label{fig:predictor_acc}\rotatebox{0}{\includegraphics[width=0.28\linewidth]{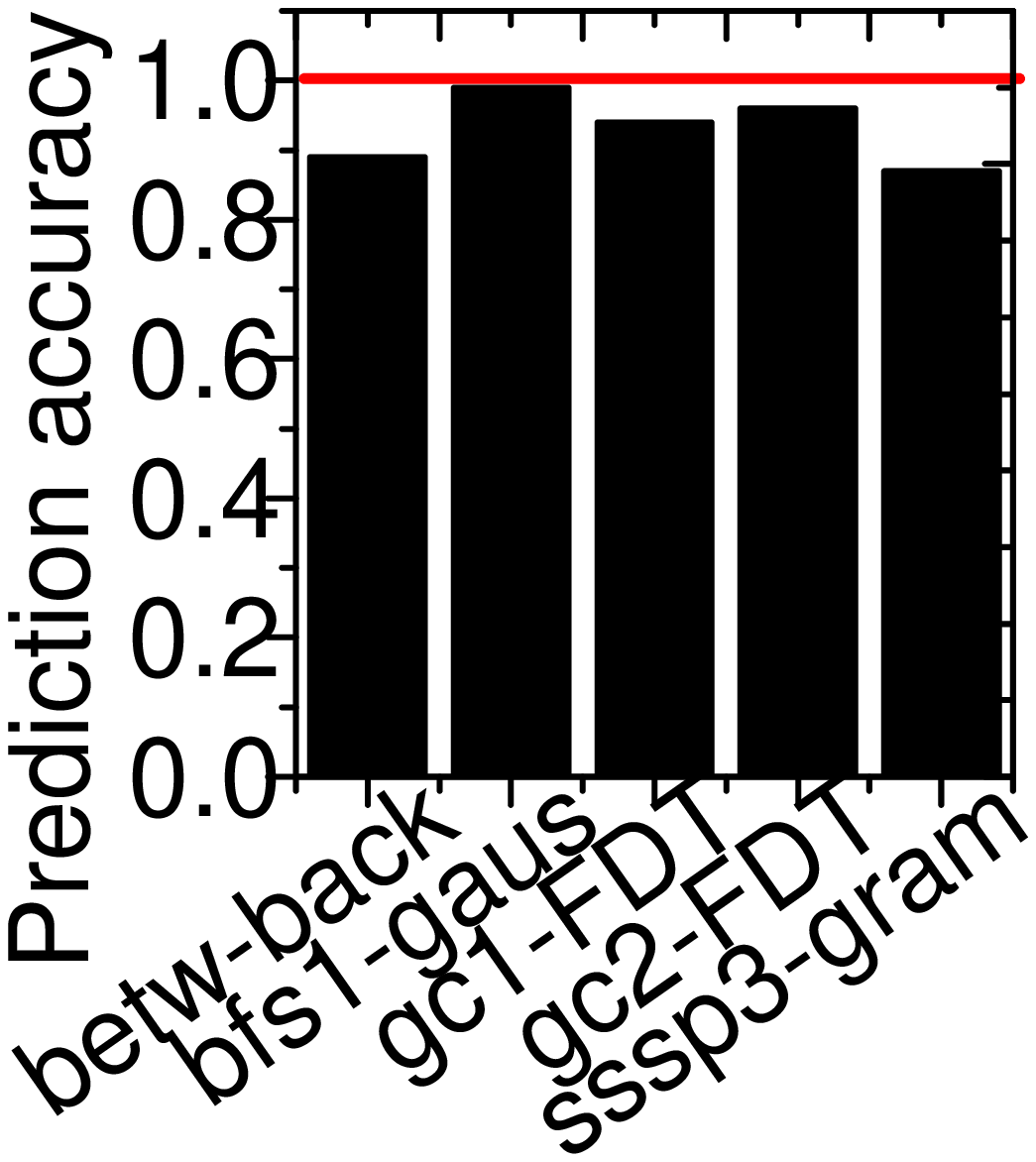}}}
\caption{\label{fig:app-acc}Scalable performance and the predictor accuracy.\vspace{-5pt}}
\vspace{-5pt}
\end{figure}

\begin{figure}
\centering
\subfloat[Varying thresholds.]{\label{fig:predictor_thres}\rotatebox{0}{\includegraphics[width=0.48\linewidth]{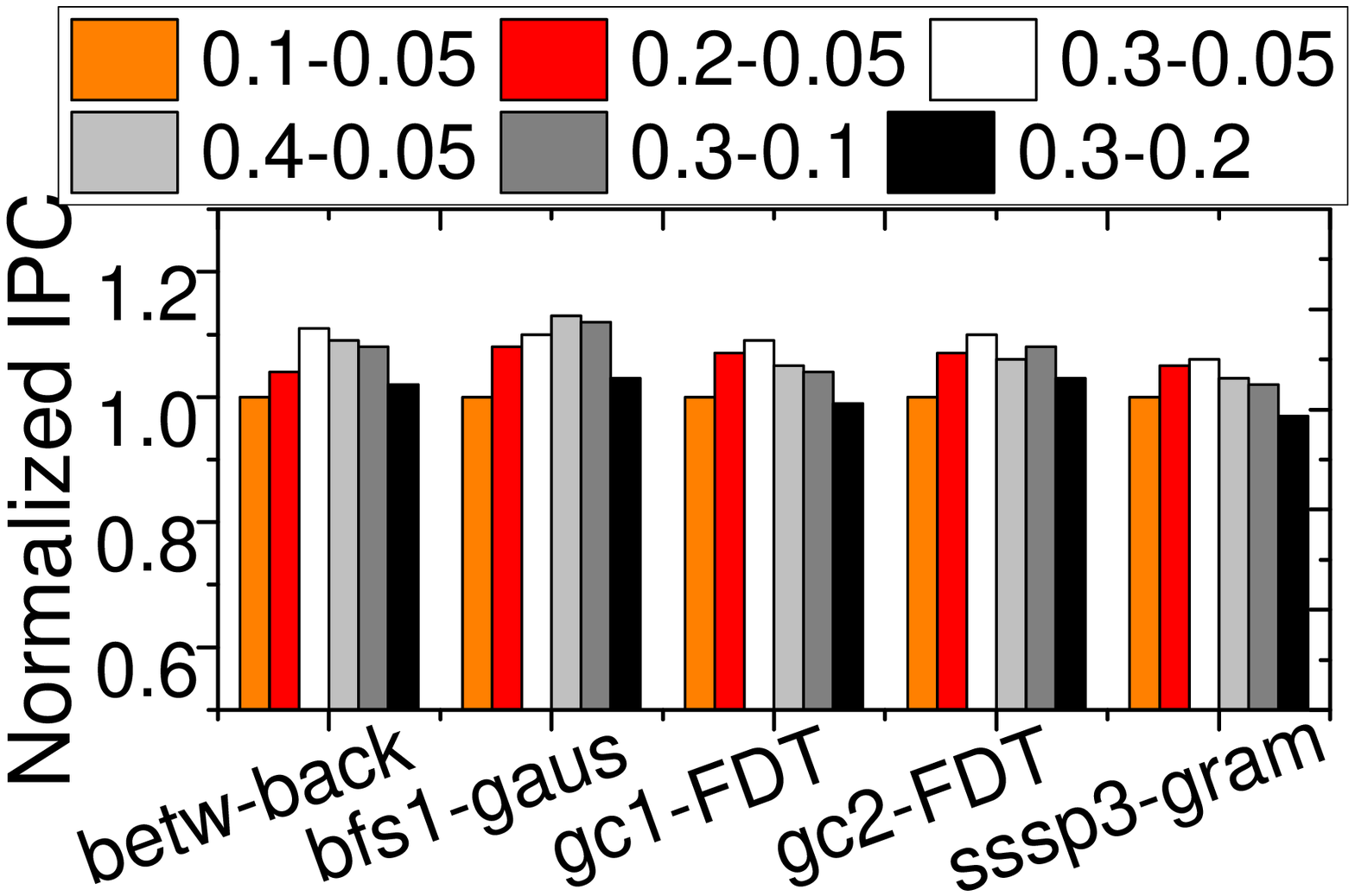}}}
\hspace{2pt}
\subfloat[Varying read prefetch methods.]{\label{fig:predictor_ipc}\rotatebox{0}{\includegraphics[width=0.48\linewidth]{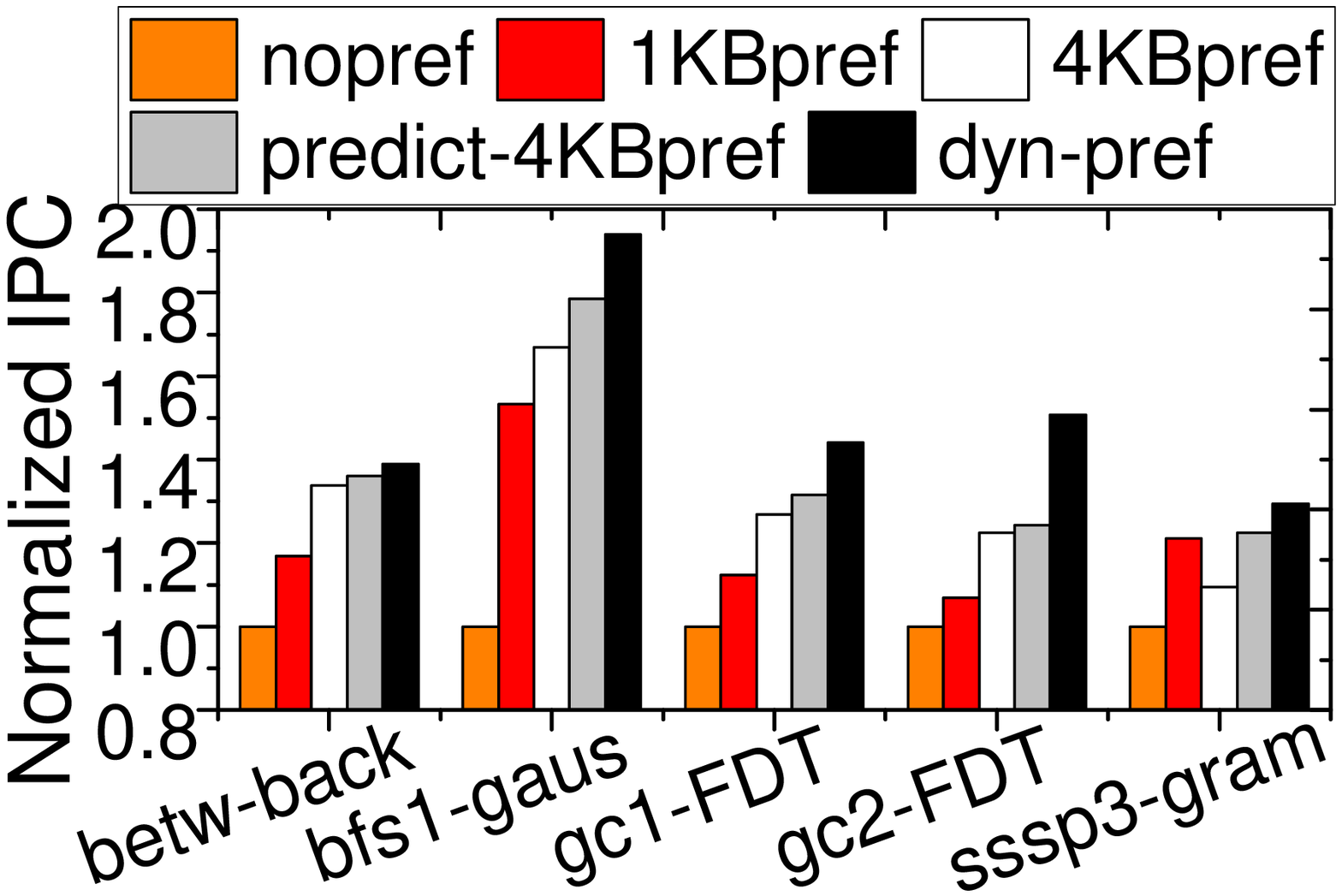}}}
\caption{\label{fig:ipc-thres}Performance analysis of prediction modules.\vspace{-5pt}}
\vspace{-15pt}
\end{figure}


\begin{figure}
\centering
\vspace{-10pt}
\subfloat[Performance of GC.]{\label{fig:gc1}\rotatebox{0}{\includegraphics[width=0.34\linewidth]{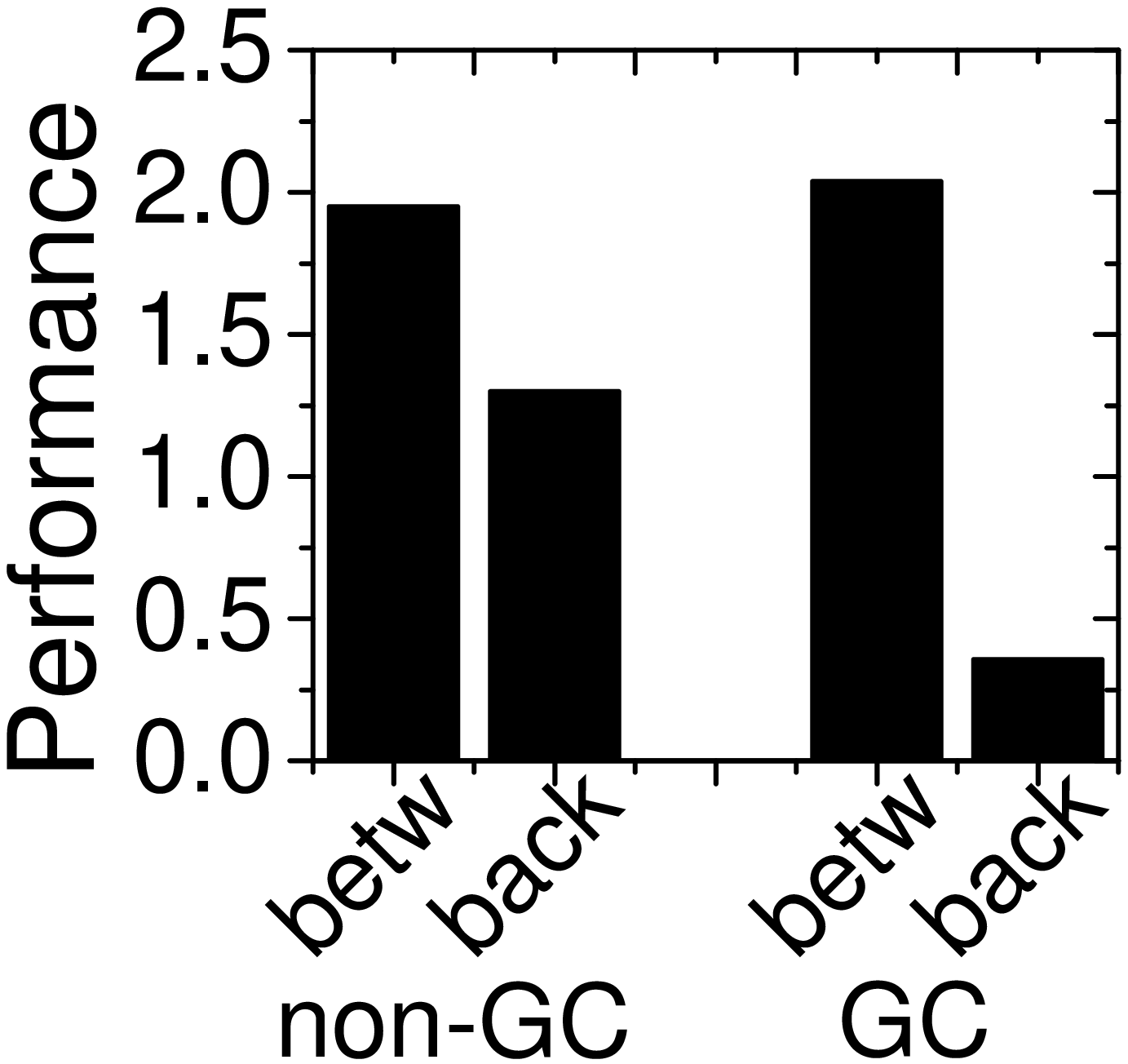}}}
\hspace{2pt}
\subfloat[Time series analysis.]{\label{fig:tsa_gc}\rotatebox{0}{\includegraphics[width=0.63\linewidth]{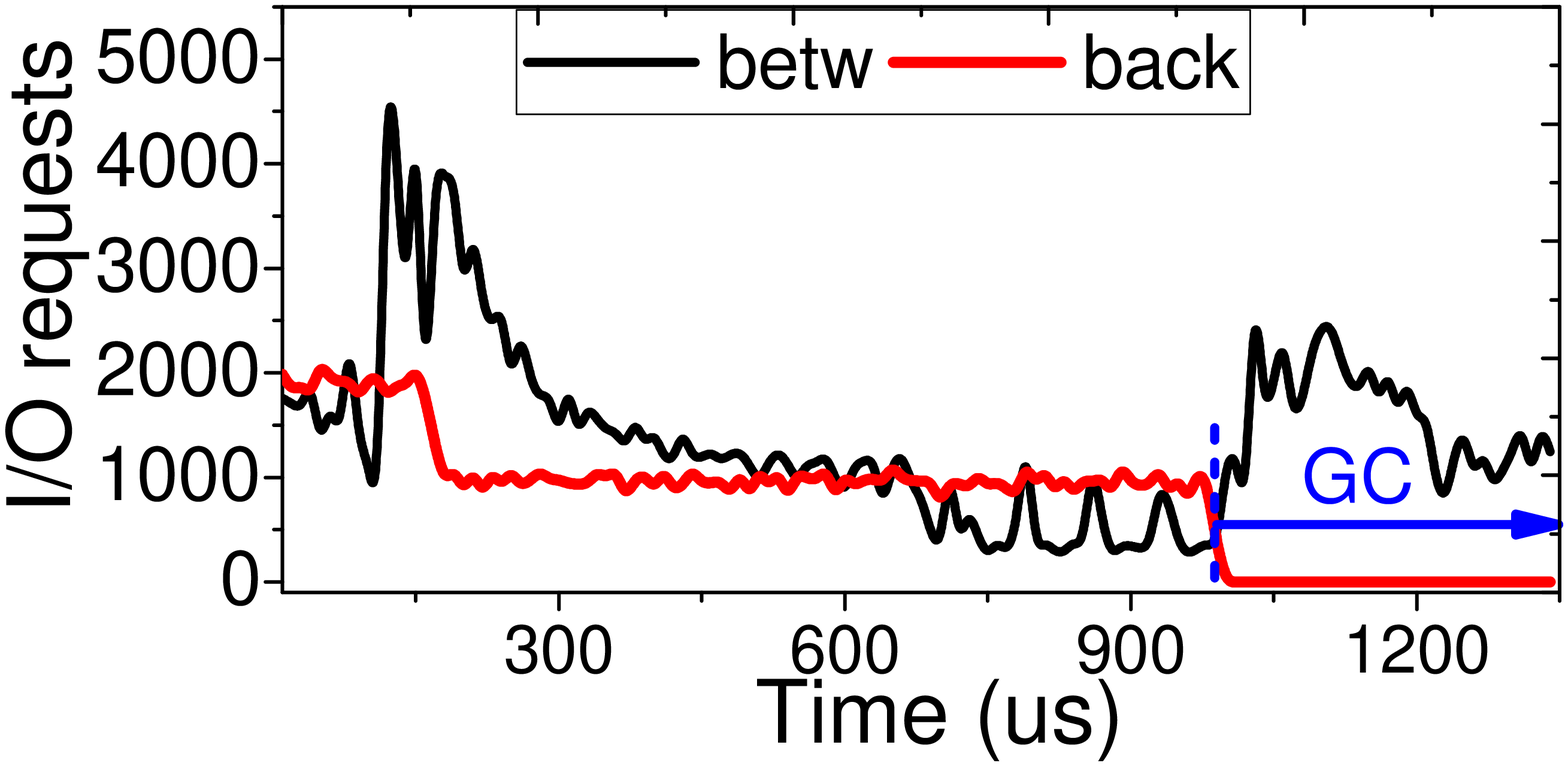}}}
\caption{\label{fig:gc}Performance analysis of garbage collection.\vspace{-5pt}}
\end{figure}

\subsection{Sensitive testing}
\label{sec:sensitive-test}
\noindent \textbf{Scalability.}
Modern GPUs support co-running multiple small-scale applications to better utilize its massive cores. We also examine performance behaviors of ZnG in co-running multiple applications, which is shown in Figure \ref{fig:multiapp1}. In the figure, we also add evaluation results of an ideal configuration for better comparison. The ideal configuration employs extra large GPU DRAM to accommodate all the data sets. We select a representative read-intensive application (\texttt{betw}) and write-intensive application (\texttt{back}) for the examination. One can observe from the figure that the performance improvement of both \texttt{Ideal} and \texttt{ZnG} is not proportional to the number of increased applications. This is because running a large number of applications in the GPU cores can generate a massive number of memory requests, which stress the underlying memory system. In practice, Amazon AWS only allows co-running upto four workloads in one GPU. Our evaluation results reveal that \texttt{ZnG} can achieve the performance improvement similar to \texttt{Ideal}, when co-running four workloads. Even if we increase the number of applications to 8, \texttt{ZnG} can still achieve 15\% and 6\% of the performance improvement, compared to \texttt{Ideal}, under the execution of read-intensive and write-intensive applications, respectively. This demonstrates the high scalability of ZnG to execute multiple applications. 

\newedit{
\noindent \textbf{Read prefetch.}
Figure \ref{fig:predictor_acc} shows the prediction accuracy of our PC-based predictor (cf. Section \ref{sec:rdopt}). Our predictor can achieve the prediction accuracy of 93\%, on average, across all the workloads, and we observe that the prediction accuracy only decreases to 87\% in the worst case.
As shown in Figure \ref{fig:predictor_thres}, we also evaluate the performance with varying high and low thresholds of our access monitor. Our access monitor can achieve the best performance if the high and low thresholds are 0.3 and 0.05, respectively.
Figure \ref{fig:predictor_ipc} compares the overall performance when disabling read prefetch (\texttt{nopref}), enabling read prefetch with 1KB/4KB data per access (\texttt{1KBpref}/\texttt{4KBpref}), and selectively prefetching 4KB data based on data locality speculation (\texttt{predict-4KBpref}) and our proposed dynamic read prefetch (\texttt{dyn-pref}). \texttt{1KBpref} and \texttt{4KBpref} can enable the read prefetch in the cases of sequential accesses, such that they outperform \texttt{nopref} by 22\% and 32\%, on average, respectively. As \texttt{predict-4KBpref} can disable the read prefetch for random accesses, it achieves better performance than \texttt{4KBpref}. Lastly, \texttt{dyn-pref} can adjust the prefetch granularity, which improves the performance by as high as 21\%, compared to \texttt{predict-4KBpref}.
}

\noindent \textbf{Garbage collection.}
Figure \ref{fig:gc1} shows the impact of garbage collection on the performance of workload \texttt{betw-back}. When the GPU helper thread reclaims the data blocks of workload \texttt{back} owing to the garbage collections, it flushes the data of \texttt{back} from L2 cache to the underlying Z-NAND and blocks all memory requests, generated by \texttt{back}, from accessing the Z-NAND. Thus, such garbage collections can unfortunately reduce the performance of \texttt{back} by 73\%, on average. However, the garbage collection procedure that we implement in ZnG does not block other applications such as \texttt{betw}. In fact, we observe that the performance of \texttt{betw} increases by 5\%, on average, due to the garbage collections of \texttt{back}. This is because the L2 cache flushes the cache lines occupied by \texttt{back} due to the garbage collection. The freed cache lines can be used for accommodating the memory requests of \texttt{betw}. Figure \ref{fig:tsa_gc} shows the time series analysis of the memory requests generated by of \texttt{betw} and \texttt{back} during their executions. Ranging from 0 us to 1108 us, the number of generated memory requests keeps decreasing in \texttt{betw} because of L2 cache competition between \texttt{betw} and \texttt{back}. As MMU blocks the memory requests during garbage collections (starting from 1108 us), the number of the memory requests, generated by \texttt{back}, decreases to 0. 

\section{Related Work and Discussion}
\label{sec:relatedwork}
\noindent \textbf{FTL integration.}
Multiple prior studies \cite{bjorling2017lightnvm, qin2019qblk, huang2015unified, abulila2019flatflash} have proposed to decouple functionalities of FTL from a flash-based SSD controller and tightly integrated the FTL into the host. \cite{bjorling2017lightnvm} proposes to allocate the flash address mapping table in the host-side main memory. Each user application in the host can create an I/O thread to update its I/O write information in the host-side flash address mapping table. However, this approach requires roughly 1GB-sized flash address mapping table for 1TB SSD. Implementing this approach in a GPU can impose a huge memory cost. In addition, the flash address translation requires involvement of the user applications rather than MMU, which decreases the performance of the user application. \cite{huang2015unified} proposes to integrate the flash address mapping table into the MMU's page table. Their approach is aimed to reduce the address translation latency of memory mapped files, rather than configuring SSD as main memory. In addition, their approach cannot reduce the size of flash address mapping table and has to place the table together with the MMU's page table in main memory. In contrast, ZnG integrates FTL in MMU, which is designed towards replacing the GPU memory with Z-NAND. Our proposed techniques reduce the mapping table size to 80KB for 1TB Z-NAND capacity, which can be placed in a MMU internal buffer. Thus, ZnG can fully eliminate the usage of DRAM. In addition, the address translation in ZnG is automated by TLB/MMU, which eliminates the FTL overhead.

\noindent \textbf{Multi-app execution.}
Modern GPUs support parallel executions of multiple applications, which can improve the utilization of GPU cores \cite{mps, armgpummu}. However, unlike traditional computer systems that employ large DRAM to store the context of tens or hundreds of applications, a GPU executes only a few number of applications simultaneously because of the limited resources (i.e., memory capacity). For example, ARM Mali-T604 can execute maximally 4 processes, as its MMU supports 4 independent address spaces \cite{armgpummu}. While the latest NVIDIA Volta GPU can submit upto 48 threads/clients into the GPU work queue, NVIDIA reports that the maximum performance improvement is limited to 7x \cite{mps}. Thus, Amazon AWS allows maximum 4 users to execute their workloads in a single GPU \cite{aws}. Our evaluation reveals that ZnG can co-run 8 graph analysis applications without significant performance degradation, which is comparable to the state-of-the-art GPUs.

\noindent \textbf{Z-NAND lifetime.}
ZnG is designed for large-scale data analysis applications such as graph analysis, which are mostly read-intensive. We employ multiple flash registers to accommodate and merge the small write requests, which can significantly reduce the number of write requests to access the underlying Z-NAND. As Z-NAND can endure much more program/erase cycles than 3D V-NAND, this guarantees Z-NAND a long lifetime. Nevertheless, we can also apply different wear-levelling algorithms in our GPU helper thread to further extend the lifetime of ZNAND. Although Optane DC PMM has much longer lifetime than Z-NAND, it exhibits a much lower memory density than Z-NAND and its accumulated bandwidth is lower than Z-NAND.

\section{Conclusion}
\label{sec:conclusion}
In this work, we propose ZnG, a new GPU-SSD architecture, which can maximize the memory capacity in a GPU, while addressing performance penalties imposed by the GPU on-board SSD. Specifically, ZnG replaces all GPU memory with a Z-NAND flash array to maximize the GPU memory capacity. It also removes an SSD controller to directly expose the accumulated bandwidth of the Z-NAND flash array to the GPU cores. Our evaluation indicates that ZnG can achieve 7.5$\times$ higher performance than prior work, on average. 

\section*{Acknowledgment}
We thank anonymous reviewers for their constructive feedback. This research is supported by NRF 2016R1C1B2015312, DOE DE-AC02-05CH 11231, KAIST Start-Up Grant (G01190015), and MemRay grant (G01190170).


\bibliographystyle{ieeetr}
\bibliography{ref}

\end{document}